\documentclass[12pt]{article}
\usepackage{amsmath,amssymb}
\usepackage{graphicx}
\usepackage{enumerate}
\usepackage{natbib}
\usepackage{times}
\usepackage{bm}
\usepackage{url} 
\usepackage{tikz}
\usetikzlibrary{arrows, decorations.markings, matrix}
\usetikzlibrary{bayesnet}
\usepackage{setspace}
\usepackage{float}
\usepackage{subfigure}
\usepackage{comment}

\newcommand{\blind}{0}

\begin{document}

\def\spacingset#1{\renewcommand{\baselinestretch}%
{#1}\small\normalsize} \spacingset{1}


\if1\blind
{
	\title{\bf  A Bayesian hierarchical model for combining multiple data sources in population size estimation}
	\author{Author 1\thanks{
			The authors gratefully acknowledge \textit{please remember to list all relevant funding sources in the unblinded version}}\hspace{.2cm}\\
		Department of YYY, University of XXX\\
		Author 2 \\
		Department of ZZZ, University of WWW}
	\maketitle
} \fi

\if0\blind
{

		    \title{\LARGE\bf A Bayesian hierarchical model for combining multiple data sources in population size estimation}

	\medskip
		\author{Jacob Lee Parsons \hspace{0.2cm}\\
		GlaxoSmithKline \vspace{0.2cm}\\
	Xiaoyue Niu \thanks{This work was supported by the National Institute of Allergy and Infectious Diseases of the National Institutes of Health under award number R01AI136664. Correspondence to: xiaoyue@psu.edu} \hspace{0.3cm} 
	Le Bao \\
	Department of Statistics, Pennsylvania State University}
	\maketitle
} \fi

\bigskip
\begin{abstract}
 To combat the HIV/AIDS pandemic effectively, targeted interventions among certain key populations play a critical role. Examples of such key populations include sex workers, people who inject drugs, and men who have sex with men. While having accurate estimates for the size of these key populations is important, any attempt to directly contact or count members of these populations is difficult. As a result, indirect methods are used to produce size estimates. Multiple approaches for estimating the size of such populations have been suggested but often give conflicting results. It is therefore necessary to have a principled way to combine and reconcile these estimates. To this end, we present a Bayesian hierarchical model for estimating the size of key populations that combines multiple estimates from different sources of information. The proposed model makes use of multiple years of data and explicitly models the systematic error in the data sources used. We use the model to estimate the size of people who inject drugs in Ukraine. We evaluate the appropriateness of the model and compare the contribution of each data source to the final estimates. 
\end{abstract}

\noindent%

\vfill

\newpage
\spacingset{1.5} 

\section{Introduction}

In order to combat the HIV/AIDS pandemic, policy makers require accurate estimates and forecasts for the number of individuals affected in each administrative unit under their care so that resources can be allocated and policy directed to the greatest effect. Many common tools for estimating and predicting the prevalence of HIV such as the Estimation and Projection Package or the Asian Epidemic Model use the size of certain key affected populations \citep{Brown2004,Ghys2004,Walker2004}. Moreover, since key affected populations are vulnerable to HIV and other health related issues, they are valuable targets for prevention programs in their own right. Some examples of these key affected populations are sex workers, people who inject drugs, and men who have sex with men. While it is important to have good estimates for the size of these key affected populations, the often marginalized nature of these groups makes any attempts to directly contact or count them difficult. These groups are therefore also referred to as the hard-to-reach population. Many techniques have been used to estimate the size of a hard-to-reach population, including but not limited to the enumeration method, the multiplier method, the capture-recapture method, the respondent-driven sampling, and the network scale-up method \citep{UNAIDS10}.

Different methods of estimation applied to the same target population at the same location will typically result in estimates that disagree to varying levels. For instance, the network scale-up method often gives higher estimates than the multiplier method \citep{Abdul2015}. Each of the methods have assumptions that must be valid for the method to result in accurate estimates. Unfortunately, the degree to which the assumptions are violated is often unknown. The deviation from the assumptions of each method result in the systematic discrepancies that are observed between resulting estimates. \cite{Abdul2015} suggest that concurrently applying multiple methods as one way of combating this problem. It is often unclear how the conflicting estimates given by various methods should be combined. Each method of estimation is subject to problems that can cause systematic errors of different magnitudes and possibly unknown directions. \cite{Okal2013} combined estimates by taking the median of the estimates under consideration, implicitly assuming that the estimates have an average of zero systematic error. \cite{Bao2015} use a Bayesian hierarchical model to combine estimates from different data sources for a region and share information between multiple regions. All of the these methods for combining estimates do not account for systematic error present in each method, and only consider estimates of a single time point. 

In the following, we produce estimates for the number of people who inject drugs in several cities of Ukraine from 2007 - 2015 by combining all publicly available estimates across multiple time points using a Bayesian hierarchical model. Our approach differs from previous methods in that we account explicitly for the systematic errors associated with the estimates and in that we consider serial correlation of data and allow the target population size to change over time. Moreover, to fill in the above-mentioned missing component in the literature of combining estimates and assessing the degree to which assumptions are violated, we provide a model to produce estimates that is accompanied by a series of model validations to quantify deviations from the model. With the question of which data sources are more important or accurate in mind, we also evaluate each data sources contribution by examining how the model's estimates and conclusions are changed by the exclusion of each data source. 

The rest of the article is organized as follows: in Section 2, we describe the available data sources in Ukraine to estimate the number of people who inject drugs. In Section 3 we propose the Bayesian hierarchical model that combines those sources. Section 4 presents the results of the model for the Ukraine data. Section 5 details the model evaluation in terms of key assumptions, goodness of fit and various source contributions for the Ukraine data. Finally, we discuss directions for further study.

\section{Available data and estimates about the size of people who inject drugs in Ukraine}

While some work has been done to estimate the size of key affected populations in Ukraine, it is not until 2009 that detailed estimates for the number of people who inject drugs in Ukraine begin to become readily available with any kind of frequency. There are two types of estimates available in Ukraine for the size of people who inject drugs at the city level: the multiplier method estimate and the network scale-up estimate. 

\subsection{Multiplier method}

One commonly used approach in estimating the hard-to-reach population is the multiplier method \citep{Johnston2013,Okal2013}. There are two pieces of data that are required to use this method: number of individuals in a subgroup of the target population $Y_k$ (e.g. number of people who inject drugs visiting a needle exchange program), and an estimate for the proportion of the target population belonging to that subgroup $p_k$ (e.g. a representative survey asking people who inject drugs whether they visited that needle exchange program). The final estimate is generated by dividing the subgroup size by the proportion estimate: 
\begin{equation}
n_k=\frac{Y_k}{p_k}.
\label{eqn:multi}
\end{equation}

The first source (subgroup) does not need to be a random sample of the population but does need to be a proper subset of the target population. The second source is assumed to be from a random sample/survey of the target population in theory, but often falls short of this in execution. It can be difficult to generate an accurate count for the size of the subgroup as directly sampling from the target population can be difficult. The resulting estimate will be inaccurate if either the subgroup size is incorrect or the proportion estimate is inaccurate. Overestimating the proportion of the target population that belong to the subgroup results in underestimating the size of the population, while underestimating the proportion results in overestimating the population. A subgroup count larger than the true size of the subgroup results in a positive bias for the estimate of size of the target population, while a smaller subgroup count results in a negative bias.

In Ukraine, surveys meant to monitor the behavior and attitudes of people who inject drugs also collect data regarding the proportion of participants that belong to certain subgroups of known size as determined by official statistics reported to the the Ukrainian government. These groups include injection drug users that stay in a drug treatment facility (DTF), those that participate in a drug treatment program (DTP), those that are hospitalized due to injection drug related causes (Hospital), those that undergo HIV rapid tests provided by non-government organizations (NGO), those that are registered to certain HIV prevention services (Prevention), those that undergo substitution maintenance therapy (SMT), and those that participate in previous behavioral surveys (Survey) \citep{Berlava2010, Berlava2012, Berlava2017}. These data are not collected every year and in no year are all of the subgroups considered. Estimates are historically generated separately for each city and then scaled up to the oblast level using an extrapolation factor.

\subsection{Network scale-up method}
 The network scale-up method (NSUM) \citep{johnsen1989estimating} is another size estimation technique and attempts to estimate the size of a hard-to-reach population using the social networks reported by individuals in a survey of the general population. The appeal of the network scale-up method lies in the simplicity of its implementation: the data could be collected by adding a few questions of the form ``How many X do you know?'' to a survey of the general population. 
 
 Let $R$ be the size of the general population, $n_k$ be the size of subpopulation $k$, $m$ be the number of respondents, $d_i$ be the network size, or degree, for person $i$, and $y_{ik}$ be the number of people that respondent $i$ reports knowing in subpopulation $k$. The basic idea behind NSUM relies on the assumption that the proportion of the general population belonging to the subpopulation is equal, on average, to the proportion of the person's network that belongs to the subpopulation, i.e.
\begin{equation*}
    \frac{y_{ik}}{d_i} = \frac{n_k}{R}
\end{equation*}

\cite{killworth1998estimation} proposed the following binomial likelihood to estimate the unknown subpopulation size $n_u$: 
\begin{equation*}
        L(n_u; \mathbf{y}, \{d_i\}) = \prod_{i = 1}^m \binom{d_i}{y_{iu}} \left(\frac{n_u}{R}\right) ^{y_{iu}} \left(1 - \frac{n_u}{R}\right)^{d_i - y_{iu}}.
\end{equation*}

The maximum likelihood estimator of $n_u$ and its standard error are the following:
\begin{eqnarray}
    \hat{n}_u^{MLE} & = & R \cdot \frac{\sum_{i = 1}^m y_{iu}}{\sum_{i = 1}^m d_i}, \\
       \mbox{SE}(\hat{n}_u^{MLE}) & = & \sqrt{\frac{R \cdot n_{u}}{\sum_{i = 1}^m d_i}}. 
    \label{eqn:mle_se}
\end{eqnarray}
 
Network scale-up method relies on three assumptions: 1. an individual is equally likely to know each other member of the population regardless of group membership; 2. each individual knows if each member of their social network is a member of the target population; and 3. all individuals have perfect recall of their complete social network. Biases may result from the violation of any of these assumptions \citep{johnsen1989estimating, johnsen1995social, killworth1998estimation, mccarty2001comparing}. Barrier effects refer to potential bias resulting from individuals having different propensities to know individuals in different groups and is typically expected to result in negatively biasing estimates. Transmissions effects refer to bias resulting from individuals not correctly identifying the membership status of individuals in their social network. When estimating the size of stigmatized populations, transmission effects may be expected to result in underestimates due to individuals hiding membership status. Recall bias refers to individuals overestimating or underestimating the number of individuals in their social network that belong to the target population as a result of being more or less likely to recall individuals in that group. More recently, several statistical models have been proposed trying to study and adjust for some of these biases \citep{zheng2006many, Maltiel2013, mccormick2010many, feehan2016generalizing}. See \cite{nsum_review} for a detailed review of the network scale-up methods.  \cite{feehan2016quantity} combine two NSUM estimates from two types of networks and obtain the weighted average as the final estimate, assuming the two estimates are independent and unbiased with the weight being their relative variances. 

\begin{table}[!h]
	\caption {Number of cities for which multiplier estimates are available for each subgroup and year combination. The last column represents the year and number of cities that have the network scale-up estimates.} \label{tab:data} 
	\begin{tabular}{lllllllll}
		    & DTF &DTP &Hospital& NGO &Prevention& SMT &Survey &Network Scale-up\\
	    2007&  0  & 0  &     14 &  0  &        0 &  0  &    0 &0\\
        2008&  0  & 0  &      0 &  0  &        0 &  0  &   0 &27\\
		2009&  0  & 0  &      0 &  0  &        0 &  14  &   0 &0\\
		2010&  0  & 0  &     12 & 26  &       21 & 23  &    0 &0\\
		2013&  0  & 0  &      0 &  0  &        0 &  0  &   27 &0\\
		2014&  0  &27  &     27 &  0  &        0 & 24  &    0 &0\\
		2015& 27  &27  &     27 & 27  &        0 & 23  &    0 &0\\
		\\
	\end{tabular}
\end{table}

\begin{figure}[!h]
	\centering
	\subfigure{\includegraphics[height=8cm]{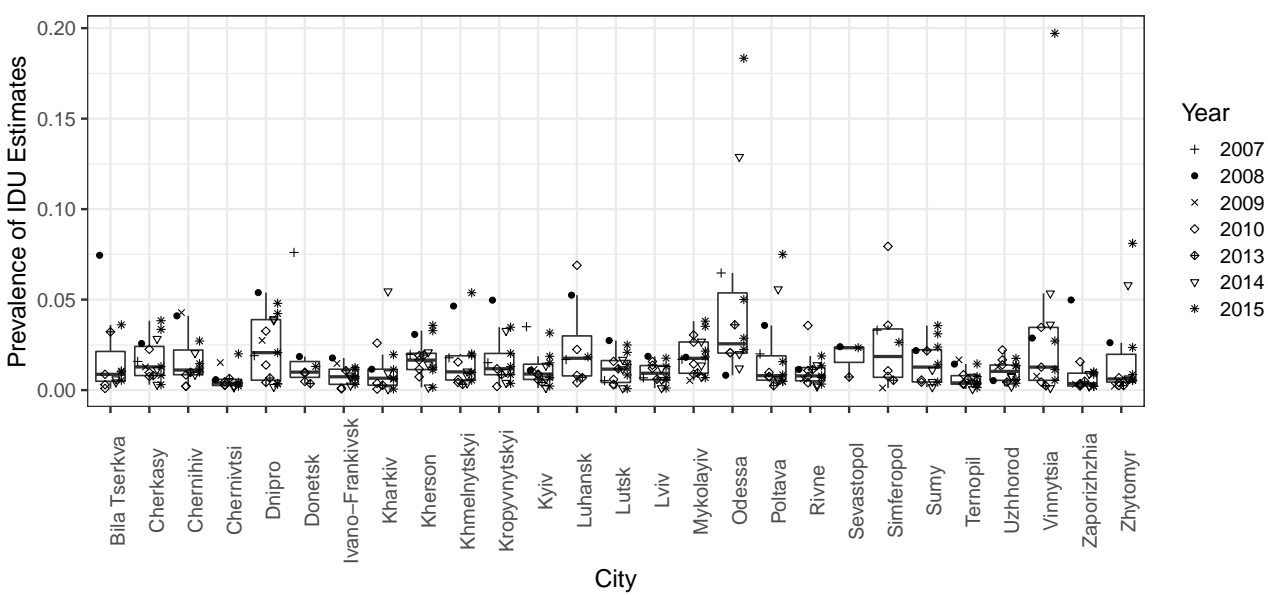}}
	\caption{Prevalence estimates for each city generated by multiplier method and network scale-up method with a symbol representing the year used to generate the estimate.  }
	\label{fig:raw_rate_estimates}
\end{figure}

In Ukraine,  27 cities have estimates in 2009 for the size of people who inject drugs that are generated using the network scale-up method (\cite{Paniotto2009}).

Table \ref{tab:data} summarizes the data availability discussed above. The number of cities by subgroup is listed for the multiplier method. Note that the year of a multiplier corresponds to the year of enrollment to the subgroup. Figure \ref{fig:raw_rate_estimates} displays the prevalence estimates generated by the multiplier method and NSUM method in each city and year. We notice that the NSUM estimates (black dots in Figure \ref{fig:raw_rate_estimates} corresponding to year 2008) are higher than most of the multiplier estimates. In addition, the heterogeneity across cities is different, with Chernivtsi's estimates quite concentrated compared to estimates from Odessa and Dnipro, for example. It should be emphasized that the cities considered in these studies are exceptional in that they all contain the administrative center for their oblast. It is therefore likely that these cities are not representative of Ukraine in general. Without further information about the population size of people who inject drugs in other cities, the determination of an extrapolation factor is a largely subjective decision. It is therefore likely that the national and oblast level estimates are less reliable than the city level estimates.

\section{A Bayesian Hierarchical Model For Size Estimation}

In our analysis we will generate estimates for the number of people who inject drugs at the city level in Ukraine from 2007 to 2015 by combining the estimates discussed in the previous section. Here our ``data" are the estimates from various sources, rather than the raw data in those sources. We denote our main estimand by $n_{it}$, the number of people who inject drugs in the $i$th city and $t$th year for each year and city combination. We first consider how to combine multiple pieces of information for a single year, and then we introduce the dynamic part of our model that utilizes multiple years of data. 

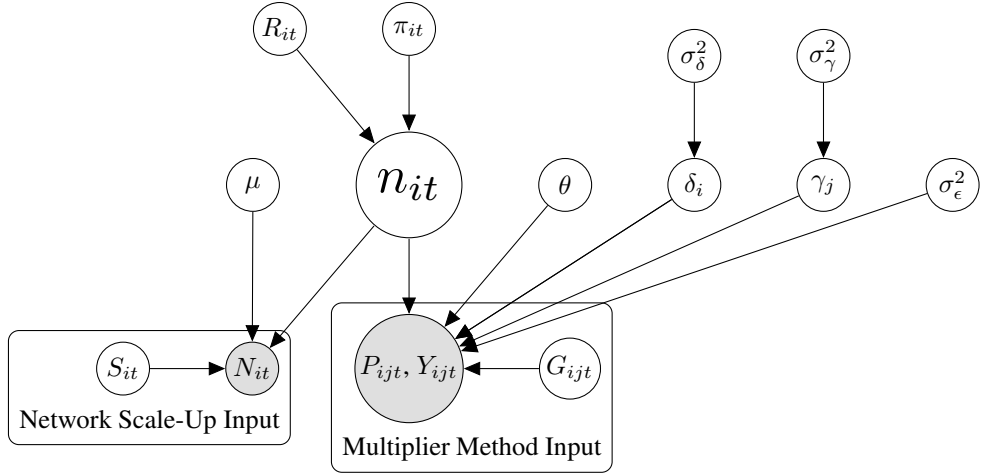
\begin{figure}[!ht]
	\centering
	\begin{tikzpicture}
	
	\node[obs] (P) {$P_{ijt}$, $Y_{ijt}$};
    \node[obs, left=of P] (N) {$N_{it}$};

	\node[latent, above=of P, scale=2] (n) {$n_{it}$};
	\node[latent, right=of n] (theta) {$\theta$};
	\node[latent, right=of theta] (delta) {$\delta_i$};
	\node[latent, right=of delta] (gamma) {$\gamma_j$};
	\node[latent, left=of n] (mu) {$\mu$};
	\node[latent, left=of N] (S) {$S_{it}$};
	\node[latent, right=of gamma] (sigmaESq) {$\sigma^2_{\epsilon}$};
	\node[latent, right=of P] (G) {$G_{ijt}$};
	
	\edge {delta} {P};
	\edge {theta} {P};
	\edge {delta} {P};
	\edge {gamma} {P};
	\edge {sigmaESq} {P};
	\edge {mu} {N};
	\edge {n} {N,P};
	\edge {S} {N};
	\edge {G} {P};
	
	\node[latent, above=of n] (pi) {$\pi_{it}$};
	\node[latent, left=of pi] (R) {$R_{it}$};
	\node[latent, above=of delta] (sigmaDeltaSq) {$\sigma^2_\delta$};
	\node[latent,above=of gamma] (sigmaGammaSq) {$\sigma^2_\gamma$};
	
	\edge {R}{n};
	\edge {pi}{n};
	\edge {sigmaGammaSq} {gamma};
	\edge {sigmaDeltaSq} {delta};

	\plate {mult} {(P)(G)} {Multiplier Method Input} ;
	\plate {nsu} {(N)(S)} {Network Scale-Up Input} ;
	
	\end{tikzpicture}
	\caption[Model Diagram for Single Year]{A diagram of the proposed model restricted to a single year. Known quantities are in capital letters and unknown quantities are in lower case or Greek letters. The random known quantities are shaded in gray. An edge pointing from a quantity to another indicates that the first quantity appears in the generating distribution of the second quantity.}
	\label{modelDiagram}
\end{figure}

For a certain year $t$, our model can be conceptualized by the diagram shown in Figure \ref{modelDiagram}. We denote all of the known quantities by capital letters and unknown quantities by lower-case or Greek letters. The data consist of two random components, the multiplier lists $Y_{ijt}$ paired with the proportion estimates $P_{ijt}$, and the NSUM estimate $N_{it}$, both shown as the shaded circles in Figure \ref{modelDiagram}. We build a statistical model around each of the two components and assume they are independent when combining them.

\subsection{Modeling the multiplier estimates}

For city $i$ and year $t$, we denote the estimated proportion of people who inject drugs falling into the $j$th subgroup by $P_{ijt}$, the sample size used to generate this estimate as $G_{ijt}$ and the total number of people who inject drugs belonging to the $j$th subgroup as $Y_{ijt}$. If the survey to estimate the proportion were a simple random sample, the multiplier estimator for $n_{it}$ is $\hat{n_{ijt}} = Y_{ijt} / P_{it}$.
However, given the hard-to-reach nature of the drug users, it is very difficult to acquire a representative sample of this target population. Therefore it is desirable to account for the bias caused by this misalignment of the sampling frame and the target population. We propose the following form:  
\begin{equation}
\text{log}(Y_{ijt} / P_{ijt}) = \text{log}(n_{it}) + \theta + \delta_i + \gamma_j + \epsilon_{ijt},
\label{eqn:p_bias2}
\end{equation}
where $\theta$ is the average bias of the multiplier method, $\delta_i$ is the city-specific bias, $\gamma_j$ is the subgroup-specific bias, and $\epsilon_{ijt}$ are zero mean independent residuals. Model (\ref{eqn:p_bias2}) allows for the bias of a survey to vary by both city and subgroup and introduces correlations for surveys within the same city and surveys for the same subgroup. We consider the choice of cities and sub-populations as sampled from a larger population and further assume the following hierarchical structure for the biases:
$$
\begin{aligned}
\delta_i \sim N(0, \sigma^2_\delta),\\
\gamma_j \sim N(0, \sigma^2_\gamma),\\
\epsilon_{ijt} \sim N(0, \sigma^2_\epsilon / G_{ijt}).
\end{aligned}
$$

The variance of $\epsilon_{ijt}$ is assumed to be inversely proportional to $G_{ijt}$, the sample size used to generate the estimate. This could be viewed as the rate of convergence if the estimate were generated by a maximum likelihood estimation and converged to asymptotic normality for large samples. This has the advantage of treating estimates generated by a large sample as more stable but not necessarily more accurate than those generated by a smaller sample.

\subsection{Modeling the network scale-up estimates}

For the NSUM data source, we denote the NSUM estimate for the target population size in city $i$ and year $t$ by $N_{it}$, and its associated standard error (also known as part of the data) as $S_{it}$. Similar to model (\ref{eqn:p_bias2}) for the  multiplier estimates, we assume the following form for the NSUM estimates:
\begin{equation}
    log(N_{it})=log(n_{it}) + \mu + e_{it},
    \label{eqn:nsum}
\end{equation}
where $\mu$ represents the average bias of the NSUM estimates. When we combine the multiplier estimates with the NSUM estimates, due to identifiability, we assume that the two sources have 0 bias on average by restricting $\mu=-\theta \sim N(0,1)$. Therefore, this bias term essentially measures the relative bias of NSUM compared to the multiplier method. Since we have only one data point to estimate this bias term, we cannot estimate the variance  but have to fix it. A prior variance at 1 implies that on the original scale the size estimate can be biased by about 7 folds ($exp(1.96)$) due to methods. In addition, we assume that the residual $e_{it} \sim N(0, \sigma^2_e)$, where $\sigma^2_e=\frac{S_{it}^2}{N_{it}^2}$ is a known quantity derived from delta method and Equation (\ref{eqn:mle_se}).

\subsection{Dynamic models}
Finally, we make use of the multiple years of data by making the framework in Figure \ref{modelDiagram} into a dynamic model. We define the prevalence of drug users in city $i$ and year $t$ as $\pi_{it}=\frac{n_{it}}{R_{it}}$, where $R_{it}$ is the reference population of city $i$ in year $t$. As seen in Table \ref{tab:data}, we have very limited information on the time dimension, which makes modeling a parametric time trend as in a regular mixed effects model unreliable. In order to make prediction, we cannot model the 7 time points as categorical levels either. Therefore, we assume the following simple random walk model with a common random drift among cities to link the prevalence across years while allowing the prevalence across cities in the same year to be correlated:
\begin{eqnarray*}
logit(\pi_{i(t+1)}) &=& logit(\pi_{it}) + \phi_{t} + \kappa_{i(t+1)},\\
\phi_{t} &\sim& N(0, \sigma_\phi^2),\\
\kappa_{it} &\sim& N(0,\sigma^2_{\pi}),\\  
logit(\pi_{i0}) &\sim& N(\mu_0,\sigma^2_0), \\
\mu_0 &\sim& N(0, 10)
\end{eqnarray*}
where $\phi_{t}$ is the national average prevalence change of year $t$ on the logit scale. $\sigma^2_{\phi}$ measures the strength of the correlation of the prevalence of different cities in the same year. This model could also be viewed as a variation of the first-order random walk model in \cite{rue05}, with the addition of the across city correlation. One key assumption here is that the missing years are missing at random so that the observed years are representative and can be used for interpolation and prediction. We do not have enough information to validate this assumption so instead we will examine the residuals later on to see whether there is any obvious violation. 

Finally, the priors of the remaining parameters are $\sigma^2_\pi,  \sigma^2_\phi, \sigma^2_\epsilon, \sigma_0^2 \sim InverseGamma(.5,.5)$. The inverse-gamma $(\alpha,\beta)$ prior is considered as a weak prior when $\alpha=\beta<1$ (\cite{gelman06}).

\section{Results}

\begin{figure}[!h]
	\centering
	\subfigure{\includegraphics[height=5.5cm]{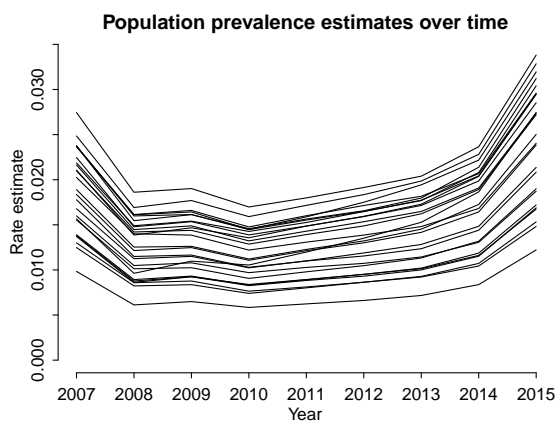}}\hfill
	\caption{Posterior mean prevalence of people who inject drugs for each year and city (A) and posterior mean size of people who inject drugs for each year and city (B). Each curve represents a city. The three cities estimated to have the fewest number of IDU are represented in dark grey while the cities with the three largest IDU size estimates are colored in black.}
	\label{fig:rate_size_estimates}
\end{figure}

\begin{figure}[!h]
	\centering
	\subfigure{\includegraphics[height=8cm]{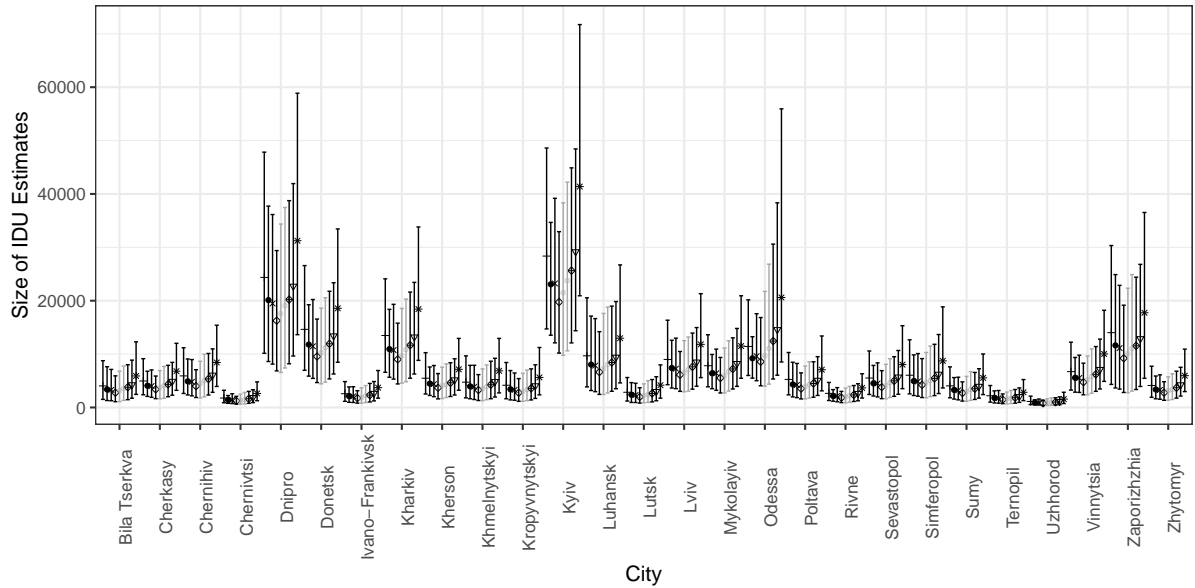}}
	\caption{Size estimates for each city generated by the proposed model, with different dots being the raw data from different years (same legend as in Figure \ref{fig:raw_rate_estimates}), and years for which data exists in black and those for which it does not in grey. }
	\label{fig:city_estimates}
\end{figure}

\begin{figure}[!ht]
	\centering
	\subfigure{\includegraphics[height=6cm]{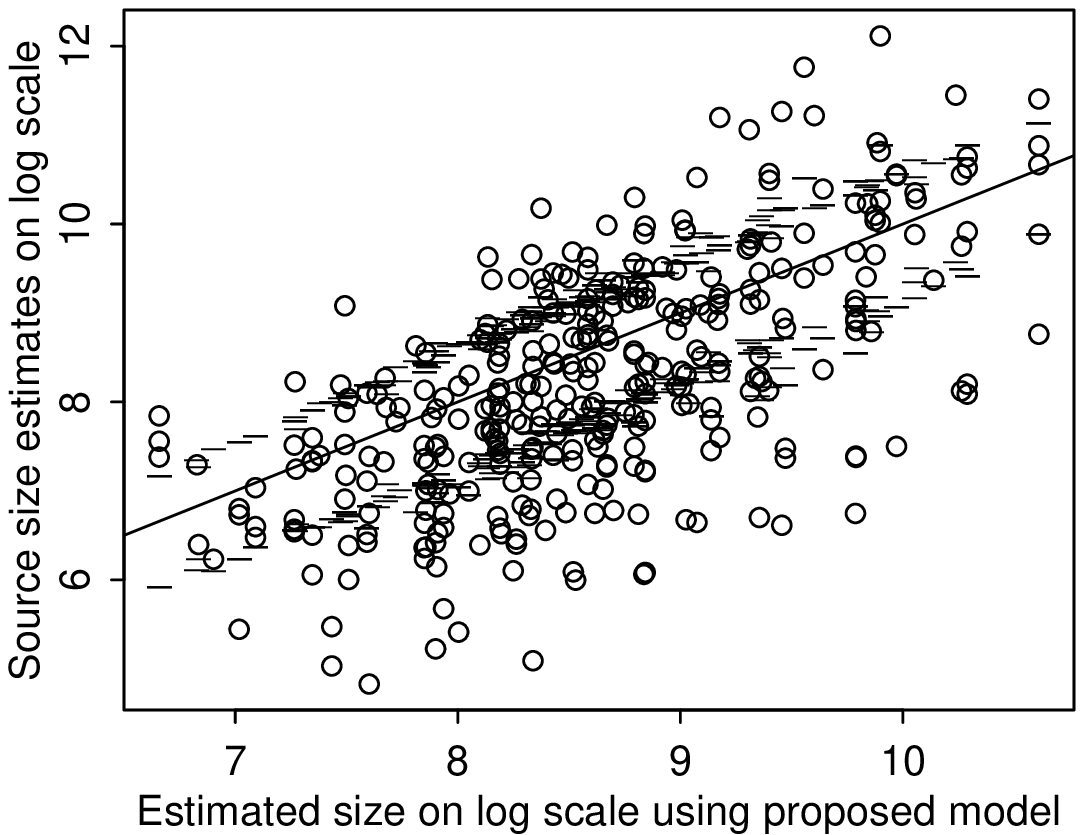}}
	\caption{Size estimates using the multiplier and network scale-up methods plotted on the y-axis with the x-axis indicating the estimate under the proposed model. $95\%$ credible interval limits for the true population size under the proposed model are indicated by short dashes, while equality between both estimates is represented by the line included in the plot. }
	\label{fig:raw_size_estimates}
\end{figure}

We apply the models described above to the Ukraine data to estimate the number of people who inject drugs in the Ukraine cities spanning the years of 2007 to 2015. We present the estimated prevalence and sizes in Figure \ref{fig:rate_size_estimates}. Most of the prevalence is between $1.0\%$ and $3.0\%$, with an increasing trend post 2010. While the cities generally have similar estimates for prevalence $\pi_{it}$, two of them have substantially higher estimates for the IDU sizes (right panel of Figure \ref{fig:rate_size_estimates}), among which Dnipro also has the highest prevalence while the Kyiv stands out due to its large population size. Figure \ref{fig:city_estimates} shows $95\%$ credible intervals for the number of IDU in each city. Note that the relatively high uncertainty of the size estimates illustrated in Figure \ref{fig:city_estimates} is reflective of the variability in the source estimates as seen in Figure  \ref{fig:raw_rate_estimates}. The width of the confidence intervals are wider when point estimates are higher, and years with no data have wider intervals compared to years with similar magnitude estimates. We also compare the model-estimated sizes with the raw data in Figure \ref{fig:raw_size_estimates}. On average, they are consistent, but the variability in the raw estimates are higher than the model-estimated ones. 

The posterior mean bias parameter $\theta$ for the multiplier method is $-.61$ and the equal-tailed $95\%$ credible interval for $\theta$ being $(-0.93, -0.27)$. Note that this interval does not include zero but exhibits a large level of uncertainty as is expected since we only have two methods to compare. This estimate is consistent with the suggestion by \citep{Abdul2015} that the NSUM results in higher estimates than the multiplier method. There are multiple potential sources of bias in both the NSUM estimates and multiplier methods that could account for this discrepancy.  For instance, bias in the multiplier method estimates is plausibly due to the inability to take a sample from the entire target population of all people who inject drugs. Some subgroups of the target population, such as individuals enrolled in a drug treatment program, would be expected to be more likely to be included in the sample. Such a positive dependence between inclusion in the chosen subgroups and in the survey could result in the bias ($\delta_i$ and $\gamma_j$). Indeed, multiple subgroups show significant group specific bias terms. We present the full results of estimated $\delta_i$ and $\gamma_j$ with their 95\% credible intervals in Figure 1 and 2 of Appendix 3.  

\section{Model Assumptions and Data Source Evaluation}

Our proposed model makes several assumptions when combining the multiple data sources, some about the structure of the model and some about the distribution of the data. In this section, we evaluate the validity of these assumptions.  In addition, we explore the contribution of each data source to the final estimates. 

\subsection{Evaluation of the structural assumptions}

We have the following assumptions about the structure of the model:
\begin{enumerate}
\item Method bias: The network scale-up and multiplier method estimates are independent. They each have a systematic bias with the average bias of the two methods being zero.
\item Constant bias: The bias of the estimates do not vary by year.
\item Multiplier bias additivity: For the multiplier estimates, the city level and subgroup level bias terms are additive.
\item Dynamic structure: The annual prevalence changes follow a random walk and the missing years are missing at random.  
\end{enumerate}
In the following subsections, we will either justify the assumptions or evaluate the effects of deviation from the assumptions. 

\subsubsection{Method bias}
The multiplier estimates and the NSUM estimates are arrived at by independently conducted surveys, thus it is reasonable to assume that they are independent. 

Figure 3 in Appendix 4.1.1 presents the results excluding the method bias term. Assuming both methods produce unbiased estimates results much more drastic annual change, with most cities having a larger than 50\% drop from 2007 to 2009 due to NSUM in 2008, and about $70\%$ of the NSUM estimates are larger than the estimated values, and $90\%$ of the estimates for the DTF and DTP subgroups are smaller than the estimates. These observations suggest that including the bias term is reasonable. In addition, assuming the average bias of the two methods are equal with opposite signs avoids favoring one method over the other. 

\subsubsection{Constant biases across years}

The availability of network scale-up estimates in multiple years or more years with  overlapping data sources could allow for further investigation of time related trends in relative bias. In the current application of our model, only one year of data exists for the network scale-up method so only varying the multiplier method bias by year will be explored here. Here we focus on allowing the average bias of the multiplier method to shift by adding a non-constant bias term $c_t$:
\begin{equation}
\text{log}(Y_{ijt} / P_{ijt}) = \text{log}(n_{it}) + \theta + c_t + \delta_i + \gamma_j + \epsilon_{ijt},
\end{equation}
where $c_t \sim N(0, \sigma_c^2)$.  

To explore the effect of the magnitude of these non-constant bias terms on estimating population size, we perform a simulation study. Details of the simulation setup can be found in Appendix 4.1.2. From the simulation we find that if we ignore the non-constant bias term in the multiplier method, on average we can still estimate the IDU size correctly (mean of the error being 0), but the uncertainty of the estimates gets much bigger as $\sigma_c$ gets bigger (increasing root-mean-squared error). Notably the effect on non-constant bias is rather muted until $\sigma_c > .4$.


\subsubsection{City and subgroup bias additivity}

We consider the additivity assumption for the subgroup and city bias terms in much the same way as we did for the constant bias in the previous section. The simulated data sets are constructed in the same way and generated from the model with an added interaction term: 
\begin{equation}
\text{log}(Y_{ijt} / P_{ijt}) = \text{log}(n_{it}) + \theta + \delta_i + \gamma_j + c_{ij} + \epsilon_{ijt},
\end{equation}
where $c_{ij} \sim N(0, \sigma_c^2)$.

Results are presented in Appendix 4.1.3. In this setting we see that the non-additive bias terms have little impact on the mean estimates. 


\subsubsection{Temporal Trend Diagnostics}
We examine the assumption we've made for the temporal trend by investigating whether there is any obvious pattern or lack of fit in the residuals, and present the results in Appendix 4.1.4. All average deviations are within reasonable ranges with no obvious temporal patterns shown. This supports that the simple random walk assumption is reasonable. 

\subsection{Evaluation of the distributional assumptions and goodness-of-fit}

\subsubsection{Distributional assumptions}
The distributions of the data being used can be checked for compatibility with the model by inspection of residuals. We inspect the behavior of the residual values based on posterior mean values of the parameters. Details can be found in Appendix 4.2.1. The residual plots all look reasonable. 

\subsubsection{Prediction via cross-validation}
While the model seems to explain the behavior of the data it used for fitting the model, it is also of interest to see how well the model predicts data that it has not yet seen. To this end, we apply a leave one site out cross validation procedure and assess the performance by the average coverage of a 95\% credible intervals for multiplier estimates and NSUM estimates, and by the correlation between observations and predictions. Details are presented in Appendix 4.2.2.

The predicted values are positively correlated with the observed values for multiplier and NSUM  estimates, with the correlations being $0.73$ and $0.66$ respectively. For the multiplier estimates, the credible intervals are somewhat conservative with $99\%$ of the $95\%$ credible intervals for the removed estimates containing the true value for the multiplier estimate. For the network scale-up estimates, the 95\% credible intervals appear well calibrated with 96\% of the credible intervals containing the true value. 

\subsubsection{Posterior predictive checks}

We use posterior predictive checks to examine how well the observed data correspond to theoretical replications of the data. We check the posterior predictive distributions of the mean of the multiplier estimates for each list and the mean of the network scale-up estimates across cities. The results are shown in Appendix 4.2.3. The observed values for these quantities are within the high probability regions for theoretical replications of these quantities under the posterior predictive distribution. 

\subsection{Data Source Contribution}

\begin{figure}[!ht]
	\centering
	\includegraphics[scale=.6]{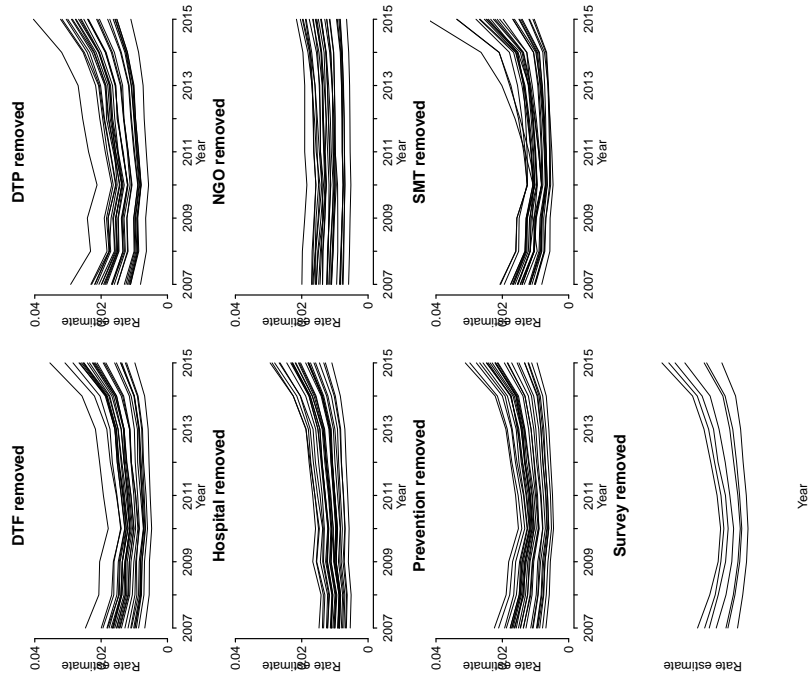}
	\caption{A. Average posterior mean prevalence of people who inject drugs across cities within each year with individual data sources removed.  B. The average width of $95\%$ credible interval for prevalence estimates across cities within a year when each data source is removed. In both plots each line corresponds to a single data source being removed and solid points indicate years for which that data source is available. }
	\label{fig:postpred2}
\end{figure}

Now we explore the contribution and impact of each subgroup on the final results. To do this we consider how the main results change when the posterior distribution is calculated without the network scale-up data and also when each subgroup is excluded from the multiplier data. Figure \ref{fig:postpred2} shows the posterior mean prevalence of people who inject drugs for each city and year combination when the posterior distribution is computed with each subgroup removed in turn. The general character of the posterior mean values remains similar in each of the plots with a couple exceptions.

First, when the posterior distribution is calculated without the Hospital group, which is the only data source for the first year, the first year has estimates that are extrapolated from the later data. This results in point estimates for the prevalence of injection drug use in each city, $\pi_{it}$, lower than what is estimated when it is included.  Removing the SMT subgroup results in estimates that are substantially lower in the years from $2010$ to $2014$, but are otherwise similar.  The exclusion of the NGO subgroup results in estimates of posterior means that are more constant across years than when it is included. The average scale of the prevalence is largely similar when excluding any single multiplier subgroup. The exclusion of the DTP subgroup leads to the largest increase in average prevalence over years. 

For those years with some data, the impact of removing a data source is similar in terms of similar widths of the credible intervals. The year 2008 has the credible intervals with the smallest width. This year has the NSU method estimates which only have a single bias term that is also shared with the multiplier method data. Years with no data available (2011 and 2012) tend to have larger and more variable impact when removing a data source, having the largest intervals and much wider range of widths across data sources.



\section{Discussion}

In summary, we propose a new hierarchical model that combines multiple sources and years of data to estimate the size of a hard-to-reach population. The proposed model allows each data source to have its own systematic bias and the population size estimate to vary over time. We evaluate the model performance in several aspects. 

As noted above, the model relies on some key assumptions. There are several reasons why those assumptions can be violated: the teams implementing the surveys may vary from year to year; there may be cultural events that change the willingness to participate among the target population; etc. A closer examination of these assumptions could be possible if there were multiple years using a combination of methods. However, our previous examination of these assumptions suggest that the proposed model is relatively robust to moderate violation of these assumptions. 

It would be desirable to have national level estimates for the number of people who inject drugs in Ukraine.  The proposed model is capable of generating national level estimates, under the assumption that the cities for which data exist are representative of the nation as a whole. Since each of these cities are the administrative center of their oblast, it is unlikely that this is true. Thus the national level of prevalence should be reinterpreted in this context to be the average prevalence among major cities in Ukraine. For more reliable national level estimates, it would be necessary to also gather data from a more representative sample of cities. Similarly, without additional information, we assume that the missing years are missing at random and homogeneity of the inclusion probabilities of those intervention programs. It is possible that these assumptions might be violated. If given further information, we could extend the model to account for the specific mechanism of the missingness of data and heterogeneous inclusion probabilities. 

The proposed model to combine multiple sources should not replace the efforts of collecting individual source of data. In fact, the large uncertainties of the estimates emphasize the importance of more systematic data collection, in consecutive years. With repeated collection of the same source, we would also be able to better study each method's systematic bias and potential reasons to further improve the method. The framework could potentially be applied to other applications of combining multiple data sources. Similar to the data fusion literature, the exact models  will depend very heavily on the specific application and data available under consideration (as they should).

\bibliographystyle{chicago}
\bibliography{Bibliography-MM-MC}

\begin{thebibliography}{}

\bibitem[\protect\citeauthoryear{Abdul-Quader, Baughman, and
  Hladik}{Abdul-Quader et~al.}{2014}]{Abdul2015}
Abdul-Quader, A.~S., A.~L. Baughman, and W.~Hladik (2014).
\newblock {Estimating the size of key populations: current status and future
  possibilities}.
\newblock {\em Current Opinion in HIV and AIDS\/}~{\em 9\/}(2), 107--114.

\bibitem[\protect\citeauthoryear{Bao, Raftery, and Reddy}{Bao
  et~al.}{2015}]{Bao2015}
Bao, L., A.~Raftery, and A.~Reddy (2015).
\newblock {Estimating the Sizes of Populations At Risk of HIV Infection From
  Multiple Data Sources Using a Bayesian Hierarchical Model}.
\newblock {\em Statistics and Its Interface\/}~{\em 8\/}(2), 125--136.

\bibitem[\protect\citeauthoryear{Berleva, Dumchev, Kasianchuk, Nikolko, Saliuk,
  Shavb, and Yaremenko}{Berleva et~al.}{2012}]{Berlava2012}
Berleva, G., K.~Dumchev, M.~Kasianchuk, M.~Nikolko, T.~Saliuk, I.~Shavb, and
  O.~Yaremenko (2012).
\newblock {Estimation of the Size of Populations Most-at-Risk for HIV Infection
  in Ukraine as of 2012}.
\newblock {\em Kyiv: International HIV/AIDS Alliance in Ukraine\/}.

\bibitem[\protect\citeauthoryear{Berleva, Dumchev, Kobyshcha, Paniotto,
  Petrenko, Saliuk, and Shvab}{Berleva et~al.}{2010}]{Berlava2010}
Berleva, G., K.~Dumchev, Y.~V. Kobyshcha, V.~I. Paniotto, T.~V. Petrenko, T.~O.
  Saliuk, and I.~A. Shvab (2010).
\newblock Analytical report based on sociological study results: Estimation of
  the size of populations most-at-risk for hiv infection in ukraine in 2009.
\newblock {\em Kyiv: International HIV/AIDS Alliance in Ukraine\/}.

\bibitem[\protect\citeauthoryear{Berleva and Sazonova}{Berleva and
  Sazonova}{2017}]{Berlava2017}
Berleva, G. and Y.~Sazonova (2017).
\newblock {Analytical report based on sociological study results{"} Estimation
  of the Size of Populations Most-at-Risk for HIV Infection in Ukraine in
  2017}.
\newblock {\em Alliance of Public Health\/}.

\bibitem[\protect\citeauthoryear{Bernard, Johnsen, Killworth, and
  Robinson}{Bernard et~al.}{1989}]{johnsen1989estimating}
Bernard, H.~R., E.~C. Johnsen, P.~D. Killworth, and S.~Robinson (1989).
\newblock Estimating the size of an average personal network and of an event
  subpopulation.
\newblock In {\em The Small World}, pp.\  159--175. Ablex Press.

\bibitem[\protect\citeauthoryear{Brown and Peerapatanapokin}{Brown and
  Peerapatanapokin}{2004}]{Brown2004}
Brown, T. and W.~Peerapatanapokin (2004).
\newblock {The Asian Epidemic Model: a process model for exploring HIV policy
  and programme alternatives in Asia}.
\newblock {\em Sexually Transmitted Infections\/}~{\em 80\/}(suppl 1),
  i19--i24.

\bibitem[\protect\citeauthoryear{Feehan and Salganik}{Feehan and
  Salganik}{2016}]{feehan2016generalizing}
Feehan, D.~M. and M.~J. Salganik (2016).
\newblock Generalizing the network scale-up method: a new estimator for the
  size of hidden populations.
\newblock {\em Sociological Methodology\/}~{\em 46\/}(1), 153--186.

\bibitem[\protect\citeauthoryear{Feehan, Umubyeyi, Mahy, Hladik, and
  Salganik}{Feehan et~al.}{2016}]{feehan2016quantity}
Feehan, D.~M., A.~Umubyeyi, M.~Mahy, W.~Hladik, and M.~J. Salganik (2016).
\newblock {Quantity Versus Quality: A Survey Experiment to Improve the Network
  Scale-up Method}.
\newblock {\em American Journal of Epidemiology\/}~{\em 183\/}(8), 747--757.

\bibitem[\protect\citeauthoryear{Gelman}{Gelman}{2006}]{gelman06}
Gelman, A. (2006).
\newblock {Prior distributions for variance parameters in hierarchical models
  (comment on article by Browne and Draper)}.
\newblock {\em Bayesian Analysis\/}~{\em 1\/}(3), 515 -- 534.

\bibitem[\protect\citeauthoryear{Ghys, Brown, Grassly, Garnett, Stanecki,
  Stover, and Walker}{Ghys et~al.}{2004}]{Ghys2004}
Ghys, P.~D., T.~Brown, N.~C. Grassly, G.~Garnett, K.~A. Stanecki, J.~Stover,
  and N.~Walker (2004).
\newblock {The UNAIDS Estimation and Projection Package: a software package to
  estimate and project national HIV epidemics}.
\newblock {\em Sexually Transmitted Infections\/}~{\em 80\/}(suppl 1), i5--i9.

\bibitem[\protect\citeauthoryear{Johnsen, Bernard, Killworth, Shelley, and
  McCarty}{Johnsen et~al.}{1995}]{johnsen1995social}
Johnsen, E.~C., H.~R. Bernard, P.~D. Killworth, G.~A. Shelley, and C.~McCarty
  (1995).
\newblock A social network approach to corroborating the number of aids/hiv+
  victims in the us.
\newblock {\em Social Networks\/}~{\em 17\/}(3-4), 167--187.

\bibitem[\protect\citeauthoryear{{Johnston }, {Prybylski Dimitri}, {Raymond H.
  Fisher}, {Mirzazadeh Ali}, {Manopaiboon Chomnad}, and {McFarland
  Willi}}{{Johnston } et~al.}{2013}]{Johnston2013}
{Johnston }, {Prybylski Dimitri}, {Raymond H. Fisher}, {Mirzazadeh Ali},
  {Manopaiboon Chomnad}, and {McFarland Willi} (2013).
\newblock {Incorporating the Service Multiplier Method in Respondent-Driven
  Sampling Surveys to Estimate the Size of Hidden and Hard-to-Reach
  Populations: Case Studies From Around the World}.
\newblock {\em Sexually Transmitted Diseases\/}~{\em 40\/}(4).

\bibitem[\protect\citeauthoryear{Killworth, McCarty, Bernard, Shelley, and
  Johnsen}{Killworth et~al.}{1998}]{killworth1998estimation}
Killworth, P.~D., C.~McCarty, H.~R. Bernard, G.~A. Shelley, and E.~C. Johnsen
  (1998).
\newblock Estimation of seroprevalence, rape, and homelessness in the united
  states using a social network approach.
\newblock {\em Evaluation review\/}~{\em 22\/}(2), 289--308.

\bibitem[\protect\citeauthoryear{Laga, Bao, and Niu}{Laga
  et~al.}{2021}]{nsum_review}
Laga, I., L.~Bao, and X.~Niu (2021).
\newblock Thirty years of the network scale-up method.
\newblock {\em Journal of the American Statistical Association\/}.
\newblock DOI: 10.1080/01621459.2021.1935267.

\bibitem[\protect\citeauthoryear{Maltiel, Raftery, and {H. McCormick}}{Maltiel
  et~al.}{2013}]{Maltiel2013}
Maltiel, R., A.~Raftery, and T.~{H. McCormick} (2013).
\newblock {Estimating Population Size Using the Network Scale Up Method}.
\newblock {\em The Annals of Applied Statistics\/}~{\em 9}.

\bibitem[\protect\citeauthoryear{McCarty, Killworth, Bernard, Johnsen, and
  Shelley}{McCarty et~al.}{2001}]{mccarty2001comparing}
McCarty, C., P.~D. Killworth, H.~R. Bernard, E.~C. Johnsen, and G.~A. Shelley
  (2001).
\newblock Comparing two methods for estimating network size.
\newblock {\em Human Organization\/}~{\em 60\/}(1), 28--39.

\bibitem[\protect\citeauthoryear{McCormick, Salganik, and Zheng}{McCormick
  et~al.}{2010}]{mccormick2010many}
McCormick, T.~H., M.~J. Salganik, and T.~Zheng (2010).
\newblock How many people do you know?: Efficiently estimating personal network
  size.
\newblock {\em Journal of the American Statistical Association\/}~{\em
  105\/}(489), 59--70.

\bibitem[\protect\citeauthoryear{Okal, Geibel, Muraguri, Musyoki, Tun, Broz,
  Kuria, Kim, Oluoch, and Raymond}{Okal et~al.}{2013}]{Okal2013}
Okal, J., S.~Geibel, N.~Muraguri, H.~Musyoki, W.~Tun, D.~Broz, D.~Kuria,
  A.~Kim, T.~Oluoch, and H.~F. Raymond (2013).
\newblock {Estimates of the size of key populations at risk for HIV infection:
  men who have sex with men, female sex workers and injecting drug users in
  Nairobi, Kenya}.
\newblock {\em Sexually Transmitted Infections\/}.

\bibitem[\protect\citeauthoryear{Paniotto, Petrenko, Kupriyanov, and
  Pakhok}{Paniotto et~al.}{2009}]{Paniotto2009}
Paniotto, V., T.~Petrenko, O.~Kupriyanov, and O.~Pakhok (2009).
\newblock {Estimating the size of populations with high risk for HIV using the
  network scale-up method}.
\newblock {\em Ukraine: Kiev International Institute of Sociology\/}.

\bibitem[\protect\citeauthoryear{Rue and Held}{Rue and Held}{2005}]{rue05}
Rue, H. and L.~Held (2005).
\newblock {\em Gaussian Markov Random Fields: Theory and Applications}.
\newblock Chapman and Hall/CRC Press.

\bibitem[\protect\citeauthoryear{{UNAIDS/WHO}}{{UNAIDS/WHO}}{2010}]{UNAIDS10}
{UNAIDS/WHO} (2010).
\newblock Guidelines on estimating the size of populations most at risk to
  {HIV}.
\newblock Technical report, UNAIDS.

\bibitem[\protect\citeauthoryear{Walker, Stover, Stanecki, Zaniewski, Grassly,
  Garcia-Calleja, and Ghys}{Walker et~al.}{2004}]{Walker2004}
Walker, N., J.~Stover, K.~Stanecki, A.~E. Zaniewski, N.~C. Grassly, J.~M.
  Garcia-Calleja, and P.~D. Ghys (2004).
\newblock {The workbook approach to making estimates and projecting future
  scenarios of HIV/AIDS in countries with low level and concentrated
  epidemics}.
\newblock {\em Sexually Transmitted Infections\/}~{\em 80\/}(suppl 1),
  i10--i13.

\bibitem[\protect\citeauthoryear{Zheng, Salganik, and Gelman}{Zheng
  et~al.}{2006}]{zheng2006many}
Zheng, T., M.~J. Salganik, and A.~Gelman (2006).
\newblock How many people do you know in prison? using overdispersion in count
  data to estimate social structure in networks.
\newblock {\em Journal of the American Statistical Association\/}~{\em
  101\/}(474), 409--423.

\end{thebibliography}

\end{document}


\def\spacingset#1{\renewcommand{\baselinestretch}%
{#1}\small\normalsize} \spacingset{1}


\if1\blind
{
	\title{\bf  A Bayesian hierarchical modeling approach to combining multiple data sources: A case study in size estimation : Supplementary Material}
	\author{Author 1\thanks{
			The authors gratefully acknowledge \textit{please remember to list all relevant funding sources in the unblinded version}}\hspace{.2cm}\\
		Department of YYY, University of XXX\\
		Author 2 \\
		Department of ZZZ, University of WWW}
	\maketitle
} \fi

\if0\blind
{

		    \title{\LARGE\bf A Bayesian hierarchical modeling approach to combining multiple data sources: A case study in size estimation : Supplementary Material}

	\medskip
		\author{Jacob Lee Parsons \hspace{0.2cm}\\
		GlaxoSmithKline \vspace{0.2cm}\\
	Xiaoyue Niu \thanks{This work was supported by the National Institute of Allergy and Infectious Diseases of the National Institutes of Health under award number R01AI136664. Correspondence to: xiaoyue@psu.edu} \hspace{0.3cm} 
	Le Bao \\
	Department of Statistics, Pennsylvania State University}
	\maketitle
} \fi

\noindent%

\spacingset{1.5} 

\section{Appendix 1: Full Conditional Distributions}

In this section, the full conditional distribution for each parameter is described. We begin by considering $\mu_0$. The full conditional distribution is such that:
$$
\begin{aligned}
  f(\mu_0 \vert \text{all other}) &\propto f(\mu_0 \vert \sigma_0) \prod_{i=1}^{I} f(logit(\pi_{i1}) \vert \mu_0, \sigma_0) \\
\end{aligned}
$$
As the $logit(\pi_{i1})$ are independent normal random variables conditional on $\mu_0$ and $\sigma_0$, the full conditional distribution is given by a normal distribution with mean
$$
\frac{\sum_{i=1}^I (logit(\pi_{i0}) )} { (\sigma^2_0/10) + I}
$$
and variance
$$
\Big( \frac{1}{10} + \frac{I}{\sigma^2_0} \Big)^{-1}.
$$
The full conditional distribution for $\pi_{i0}$ is given by 
$$
\begin{aligned}
  f(\pi_{i0} \vert \text{all other}) &\propto f(\pi_{i0} \vert \alpha_0, \beta_0) f(\pi_{i1} \vert \pi_{i0}, \sigma_\pi, \phi_1) f(log(N_{i0}) \vert \pi_{i0}, \mu) \prod \limits_{j=1}^J f( M_{ij0} \vert \pi_{i0}, \sigma_P, \theta, \gamma_j, \delta_i   )  \\
\end{aligned}
$$
where we use $M_{ijt}$ to be the log of the multiplier estimate $Y_{ijt} / P_{ijt}$ and
$$
\begin{aligned}
f(\pi_{i0} \vert \mu_0, \sigma_0) &=  \sqrt{1/ 2 \pi \sigma^2_0}  \cdot  exp \Big [-\frac{1}{2\sigma_0^2} \sum_{i=1}^I  \big( logit(\pi_{i0}) - \mu_0   \big)^2  \Big] ,\\
f(M_{ijt} \vert \pi_{ijt}, \theta, \delta_i, \gamma_j) &= \sqrt{G_{ijt}/2 \pi \sigma^2_p}  exp \Big [-\frac{G_{ijt}}{2\sigma_P^2} \big( M_{ijt} - log(\pi_{it} R_{it}) - \theta - \delta_i - \gamma_j   \big)^2  \Big] ,\\
f(log(N_{it}) \vert \pi_{ijt}, \mu) &= (2 \pi \pi_{it} D_{it})^{-1/2} \sigma^2_p exp \Big [ (2 \pi_{it} D_{it})^{-1/2}  \big( log(N_{it}) - log(\pi_{it} R_{it}) - \mu   \big)^2  \Big] ,\\
\text{and \space} f(\pi_{i1} \vert \pi_{i0}, \sigma_\pi, \phi_1) &= \frac{1}{\sqrt{2\pi\sigma_\pi^2}}  exp\Big(-\frac{1}{2\sigma_\pi^2}[logit(\pi_{i1}) - logit(\pi_{i0}) - \phi_1 ]^2 \Big).
\end{aligned}
$$
For $0< t < T$, the full conditional distribution for $\pi_{it}$ is given by 
$$
\begin{aligned}
  f(\pi_{it} \vert \text{all other}) &\propto  f(N_{it} \vert \pi_{it}, R_{it}) f(\pi_{it} \vert \pi_{i(t-1)}, \sigma_\pi, \phi_t) f(\pi_{i(t+1)} \vert \pi_{it}, \sigma_\pi, \phi_{t+1})  \\ 
   & \quad \quad \quad \times \prod \limits_{j=1}^J f( M_{ijt} \vert \pi_{it}, \sigma_P, \theta, \gamma_j, \delta_i   )  \\
\end{aligned}
$$
where in addition to the functions previously introduced,
$$
\begin{aligned}
f(\pi_{i(t+1)} \vert \pi_{it}, \sigma_\pi, \phi_{t+1} &= \frac{1}{\sqrt{2\pi\sigma_\pi^2}} exp\Big(-\frac{1}{2\sigma_\pi^2}[logit(\pi_{i(t+1)}) - logit(\pi_{it}) - \phi_{(t+1)} ]^2 \Big).
\end{aligned}
$$
Similarly, for the final year when $t=T$, 
$$
\begin{aligned}
  f(\pi_{it} \vert \text{all other}) &\propto  f(N_{it} \vert \pi_{it}, R_{it}) f(\pi_{it} \vert \pi_{i(t-1)}, \sigma_\pi, \phi_t)  \prod \limits_{j=1}^J f( M_{ijt} \vert \pi_{it}, \sigma_P, \theta, \gamma_j, \delta_i   )  \\
\end{aligned}
$$
Many parameters use a normal prior as a conjugate distribution for the mean of a normal distribution yielding simple full conditional distributions. For instance, $\theta$ has a normal full conditional distribution with mean
$$
\begin{aligned}
&V \Big( \sum \limits_{i=1}^I \sum \limits_{t=1}^T (\pi_{it} D_{it})^{-1}[ log(N_{it}) - log(R_{it}) \pi_{it} ]  \\
 &\quad \quad \quad \quad \quad \quad + \sum \limits_{i=1}^I \sum \limits_{j=1}^J \sum \limits_{t=1}^T G_{ijt} \sigma_\epsilon^{-2} ( log(M_{ijt}) - log(n_{it}) - \delta_i - \gamma_j )  \Big)
\end{aligned}
$$
and variance
$$
V = \Big(1 + \sum \limits_{i=1}^I \sum \limits_{t=1}^T (\pi_{it} D_{it})^{-1} + \frac{\sum_{i=1}^I \sum_{j=1}^J \sum_{t=1}^T G_{ijt}}{\sigma_\epsilon^2}   \Big)^{-1}
$$
The full conditional distribution for $\gamma_j$ is normal with mean
$$
\frac{\sum \limits_{i=1}^I \sum \limits_{t=1}^T G_{ijt} ( log(M_{ijt}) - log(n_{it}) - \theta - \delta_i )}{\frac{\sigma_\epsilon^2}{\sigma_\gamma^2} + \sum \limits_{i=1}^I \sum \limits_{t=1}^T G_{ijt} } 
$$
and variance
$$
\Big({\frac{1}{\sigma_\gamma^2} + \frac{\sum_{i=1}^I \sum_{t=1}^T G_{ijt}}{\sigma_\epsilon^2} }\Big)^{-1}.
$$
The full conditional distribution for $\delta_i$ is normal with mean
$$
\frac{ \sum \limits_{j=1}^J \sum \limits_{t=1}^T G_{ijt}( log(M_{ijt}) - log(n_{it}) - \theta - \gamma_j) }{\frac{\sigma_\epsilon^2}{\sigma_\delta^2} + \sum \limits_{j=1}^J \sum \limits_{t=1}^T G_{ijt} } 
$$
and variance
$$
\Big({\frac{1}{\sigma_\delta^2} + \frac{\sum_{j=1}^J \sum_{t=1}^T G_{ijt}}{\sigma_\epsilon^2} }\Big)^{-1}.
$$
The full conditional distribution for $\phi_t$ is normal with mean
$$
\frac{\sum \limits_{i=1}^I logit(\pi_{it}) - logit(\pi_{i(t-1)}) }{\frac{\sigma_\pi^2}{\sigma_\phi^2} + I } 
$$
and variance
$$
\frac{1}{\frac{1}{\sigma_\phi^2} + \frac{I}{\sigma_\pi^2} }.
$$
The inverse-gamma prior for $\sigma_\pi^2$ is a conjugate prior for the variance in the normal distribution on the prevalence terms. Thus, the full conditional distribution for $\sigma_\pi^2$ is inverse gamma with shape parameter $.5 + \frac{IT}{2}$ and scale parameter given by
$$
.5 + .5 \sum \limits_{i=1}^I\sum \limits_{t=1}^T (logit(\pi_{it}) - logit(\pi_{i(t-1)}) - \phi_t)^2.
$$
The other scale parameters are handled in the same way. The full conditional distribution for $\sigma_\gamma^2$ is inverse-gamma with shape parameter $1 + \frac{J}{2}$ and scale parameter given by
$$
.001 + .5 \sum \limits_{j=1}^J  \gamma_j^2.
$$
The full conditional distribution for $\sigma_\delta^2$ is inverse-gamma with shape parameter $1+ \frac{I}{2}$ and scale parameter given by
$$
.001 + .5 \sum \limits_{i=1}^I  \delta_i^2.
$$
The full conditional distribution for $\sigma_\epsilon^2$ is inverse-gamma with shape parameter
$$
1 + \frac{\sum_{i=1}^I \sum_{j=1}^J \sum_{t=1}^T G_{ijt}}{2}
$$
and scale parameter given by
$$
.001 + .5 \sum \limits_{i=1}^I \sum \limits_{j=1}^J \sum \limits_{t=1}^T G_{ijt} (log(M_{ijt}) - log(n_{it}) - \theta - \delta_i - \gamma_j)^2.
$$
THe full conditional distribution for $\sigma_0^2$ is also inverse-gamma with shape parameter
$$
.5 + I/2
$$
and scale parameter
$$
.5 + \sum \limits_{i=1}^I  (logit(\pi_{i0}) - \mu_0)^2 /2.
$$

\section{Appendix 2: Markov Chain Monte Carlo Algorithm and Proposal Distributions}

The Markov chain Monte Carlo algorithm used is a straight forward Metropolis-Hastings algorithm that uses Gibbs updates whenever available (see Appendix 1). In what follows we detail the proposal distributions used for the remaining variables.

The proposal distribution for $log(\pi_{it})^{(m)}$ is normal with mean  $log(\pi_{it})^{(m-1)}$ and a standard deviation that should be chosen based on the particular situation. For the main analysis, a standard deviation of .4 was chosen. 

Code for the algorithm is available online at online at:

\href{https://github.com/JacobLParsons/SizeEstimationModel}{https://github.com/JacobLParsons/SizeEstimationModel}.

\section{Appendix 3: Site and Subgroup Specific Bias Estimates}

\begin{figure}[!h]
	\centering
	\includegraphics[height=7.5cm]{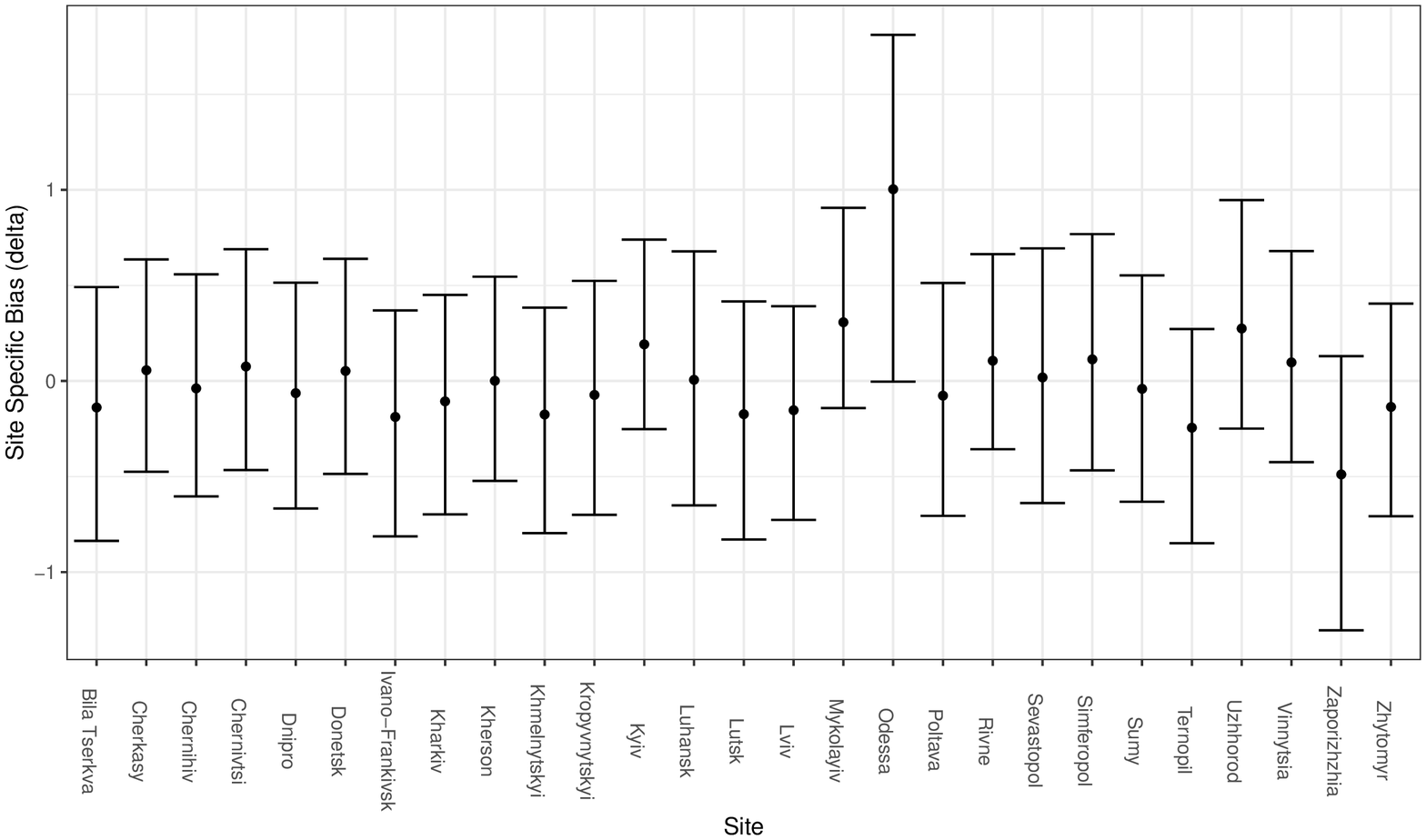}
	\caption{Site specific bias estimates for the multiplier method in each city with $95\%$ credible intervals.  }
	\label{fig:delta}
\end{figure}

\begin{figure}[!h]
	\centering
	\includegraphics[height=7.5cm]{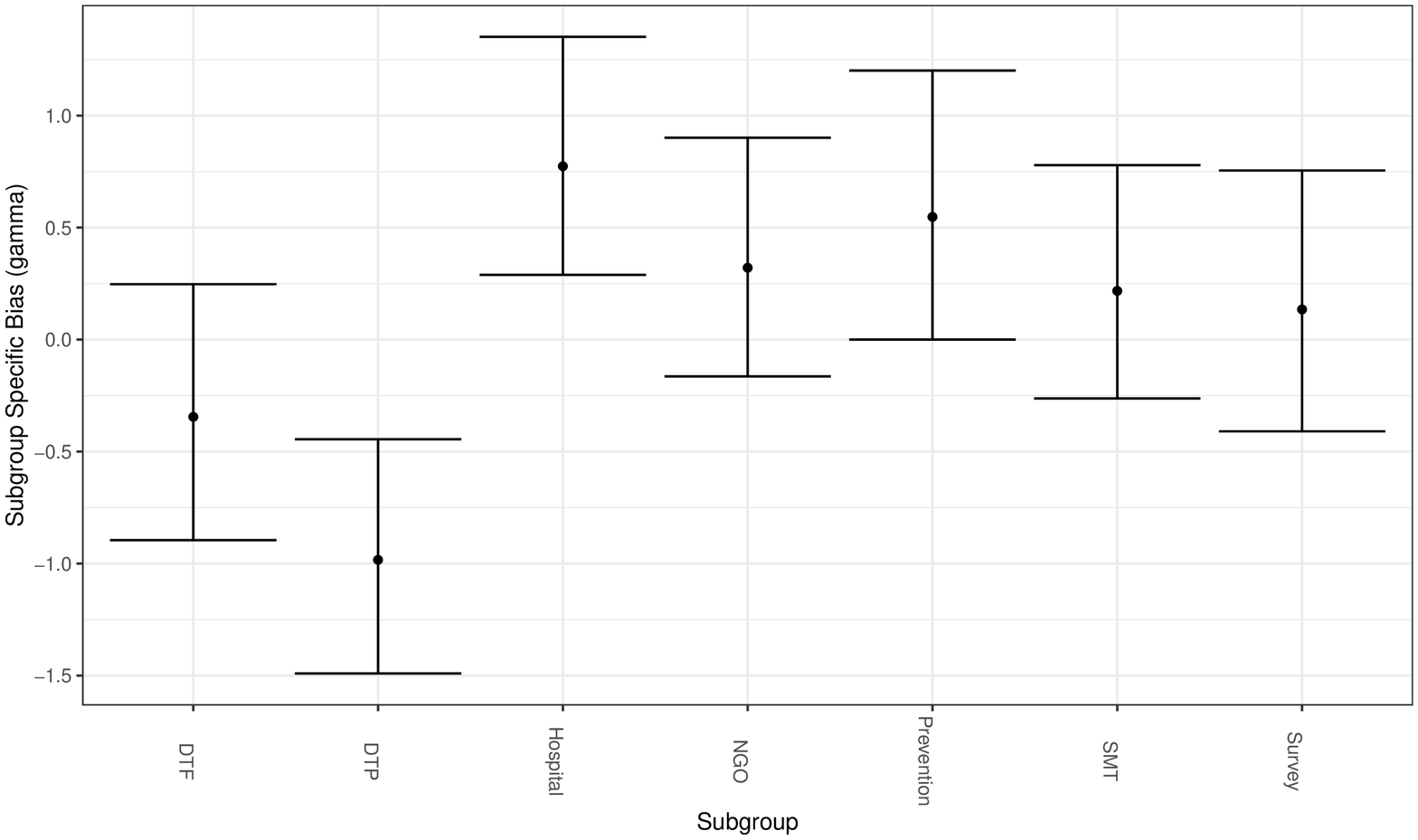}
	\caption{Subgroup specific bias estimates for the multiplier method with $95\%$ credible intervals.  }
	\label{fig:gamma}
\end{figure}

We present the estimated city-specific bias and list-specific bias with their credible intervals in Figures \ref{fig:delta} and \ref{fig:gamma}. The cities do not seem to vary too much while many of the list effects are significant. 

\newpage

\section{Appendix 4: Model Evaluation}

\subsection{Evaluation of the structural assumptions}

We have the following assumptions about the structure of the model:
\begin{enumerate}
\item Method bias: The network scale-up and multiplier method estimates are independent. They each have a systematic bias with the average bias of the two methods being zero.
\item Constant bias: The bias of the estimates do not vary by year.
\item Multiplier bias additivity: For the multiplier estimates, the city level and subgroup level bias terms are additive.
\item Dynamic structure: The annual prevalence changes follow a random walk and the missing years are missing at random.  
\end{enumerate}
In the following subsections, we will either justify the assumptions or evaluate the effects of deviation from the assumptions. 

\subsubsection{Method bias}
The multiplier estimates and the NSUM estimates are arrived at by independently conducted surveys, thus it is reasonable to assume that they are independent. 

\begin{figure}
	\centering
	\subfigure{\includegraphics[height=5.5cm]{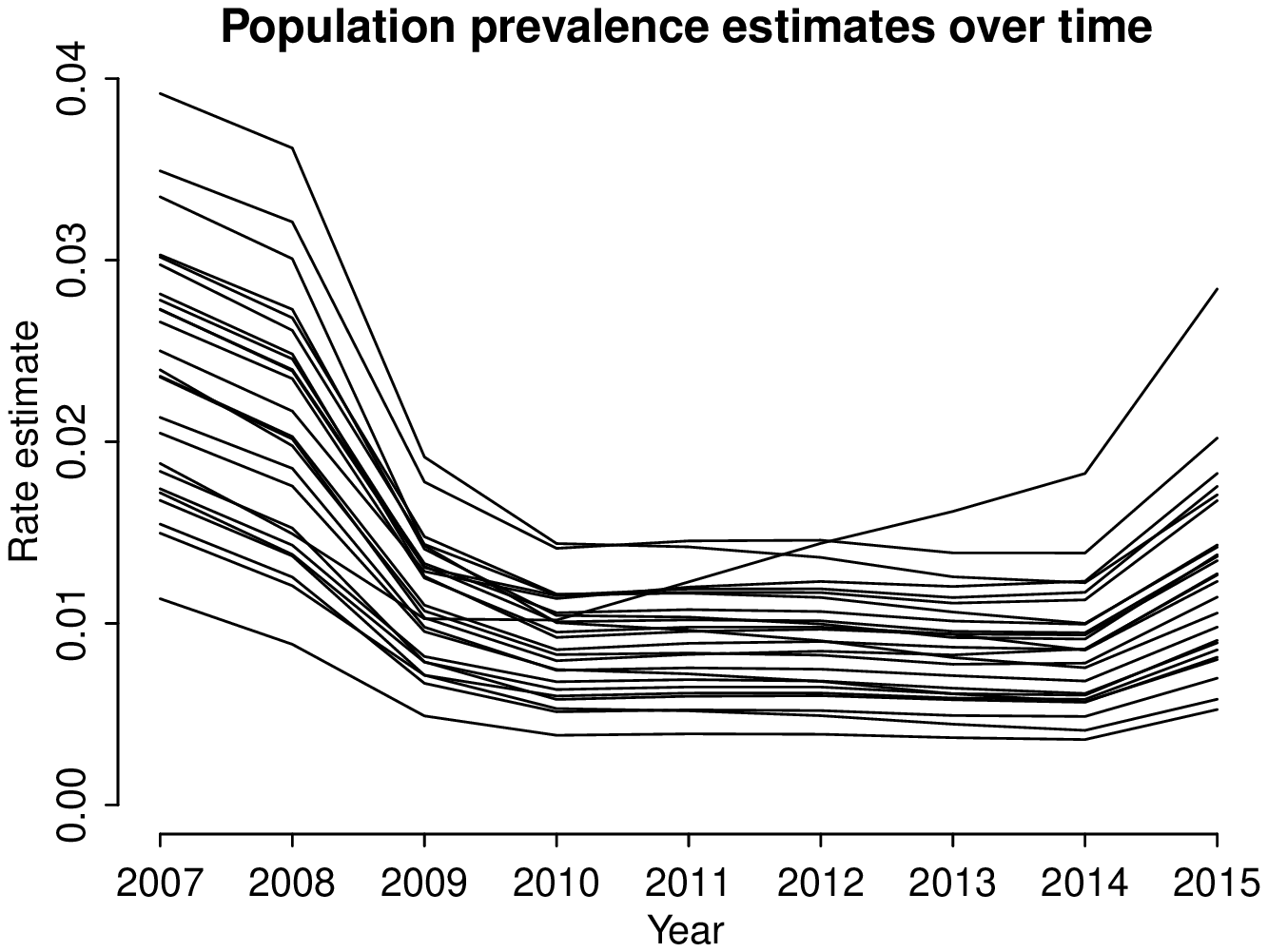}}
	\subfigure{\includegraphics[height=5.5cm]{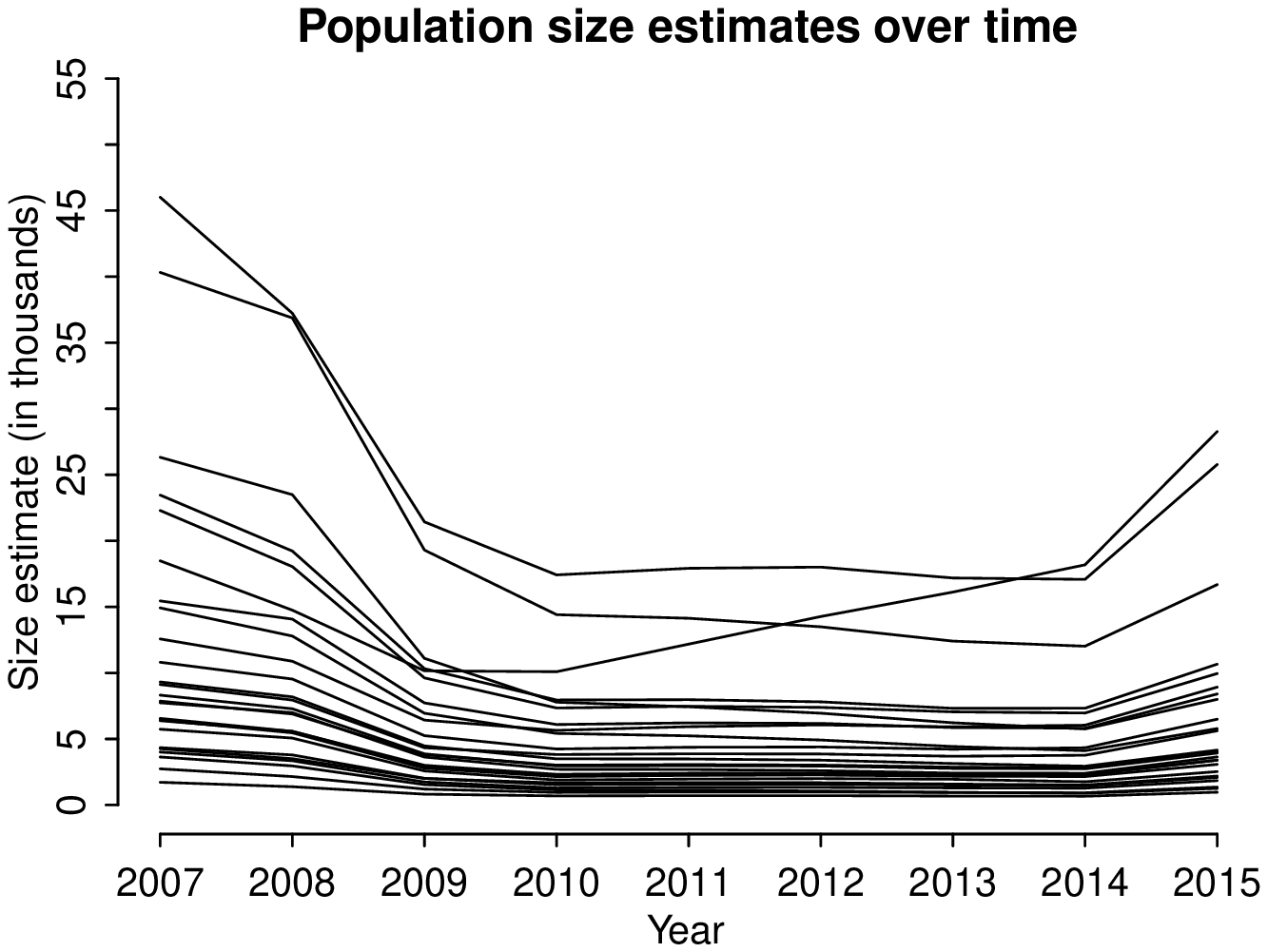}}
	\caption{Posterior mean prevalence of people who inject drugs for each year and city (left) and posterior mean size of people who inject drugs for each year and city (right) under the assumption of no bias. }
	\label{fig:rate_size_estimates_nb}
\end{figure}

We observe that the two methods give systematically different results, with multiplier estimates higher than the NSUM estimates. Given that the NSUM estimates are in a year (2008) when no multiplier estimates are available, assuming both methods produce unbiased estimates of the true sizes would force annual change to be more drastic. Figure \ref{fig:rate_size_estimates_nb} shows the mean posterior prevalence and size for each city and year when the bias terms are excluded from the model.

 We can see a big drop of IDU sizes from 2007 to 2009, with most cities having a larger than 50\% drop over the course of two years. Moreover, about $70\%$ of the NSUM estimates are larger than the estimated values, and $90\%$ of the estimates for the DTF and DTP subgroups are smaller than the estimates. These observations suggest that including the bias term is reasonable. In addition, assuming the average bias of the two methods are equal with opposite signs avoids favoring one method over the other. 

\subsubsection{Constant biases across years}

The availability of network scale-up estimates in multiple years or more years with  overlapping data sources could allow for further investigation of time related trends in relative bias. In the current application of our model, only one year of data exists for the network scale-up method so only varying the multiplier method bias by year will be explored here. Here we focus on allowing the average bias of the multiplier method to shift according to the following model:
\begin{equation}
\text{log}(Y_{ijt} / P_{ijt}) = \text{log}(n_{it}) + \theta + c_t + \delta_i + \gamma_j + \epsilon_{ijt},
\end{equation}
where $c_t \sim N(0, \sigma_c^2)$. We shall refer to $c_t$ as the non-constant bias term. 

To explore the effect of the magnitude of these non-constant bias terms on estimating population size, we perform a simulation study. In the simulation study we simulate 400 data sets. $R_{it}$ is generated uniformly between 20,000 and 400,000. $\pi_{it}$ is generated according to the dynamic model in Section 3.3 with $\mu_0\sim$ N(logit(0.1), 0.5) to mimic the real data. We sample all of the variance parameters uniformly between 0 and 1, except for $\sigma^2_\epsilon$.  $\sigma^2_\epsilon$ is sampled uniformly between 200 and 500, so the uncertainty ranges from smaller than what we estimate to about twice as large. Note that $\sigma^2_\epsilon$ is much larger than the other variances because the errors are generated with the scaled variance: $\epsilon_{ijt} \sim N(0, \sigma^2_\epsilon / G_{ijt})$. The simulated data sets include 20 cities, 4 years, and 4 multiplier sub-groups where the second year contains only NSU data and the remaining three only multiplier method estimates. 
\begin{figure}[!ht]
	\centering
		\includegraphics[scale = .5, angle=270]{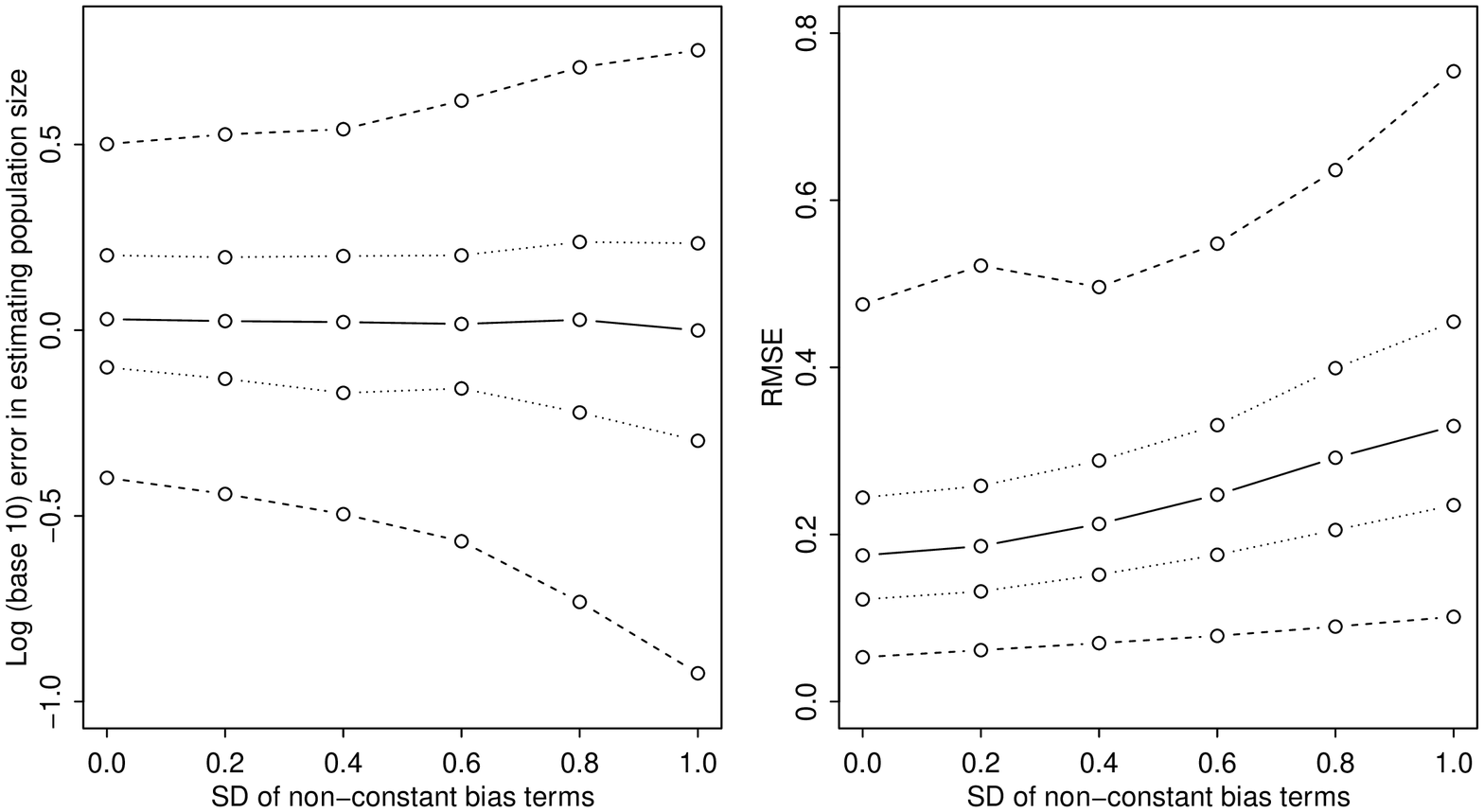}
	\caption{The left panel shows the biases in estimating the population size for a city on the $log_{10}$ transformed scale plotted against the standard deviation $\sigma_c$ of the non-constant bias terms $c_{t}$. The solid middle line indicates the mean error in estimation across simulated data sets and cities, the middle dotted lines indicate bounds containing $50\%$ of the simulated average error across all cities and year combinations in each data set. The dashed lines indicate the middle $95\%$ errors in estimating population size for each city and year combination across all data sets. The right panel plots the mean root-mean square error in estimating the population size of all cities and years in each simulated data set with the solid line indicating the mean root mean square error, the dotted lines indicating the middle $50\%$ root mean square errors, and the dashed lines the middle $95\%$. }
		\label{fig:ConstantBias}
\end{figure}
Figure \ref{fig:ConstantBias} shows the distributions of the errors (showing direction of the bias) and root-mean square errors (showing variance of the bias) observed in the simulations. It can be seen that the spread of the average bias across all cities and time points doesn't increase dramatically, but the spread of the individual errors does notably increase. This is reflected in a fairly substantial increase in the root-mean square error for estimates in the right panel of Figure \ref{fig:ConstantBias}. Notably the effect on non-constant bias is rather muted until $\sigma_c > .4$.


\subsubsection{City and subgroup bias additivity}

We consider the additivity assumption for the subgroup and city bias terms in much the same way as we did for the constant bias in the previous section. The simulated data sets are constructed in the same way and generated from the model below: 
\begin{equation}
\text{log}(Y_{ijt} / P_{ijt}) = \text{log}(n_{it}) + \theta + \delta_i + \gamma_j + c_{ij} + \epsilon_{ijt},
\end{equation}
where $c_{ij} \sim N(0, \sigma_c^2)$. 

\begin{figure}[!ht]
	\centering
		\includegraphics[scale = .5, angle=270]{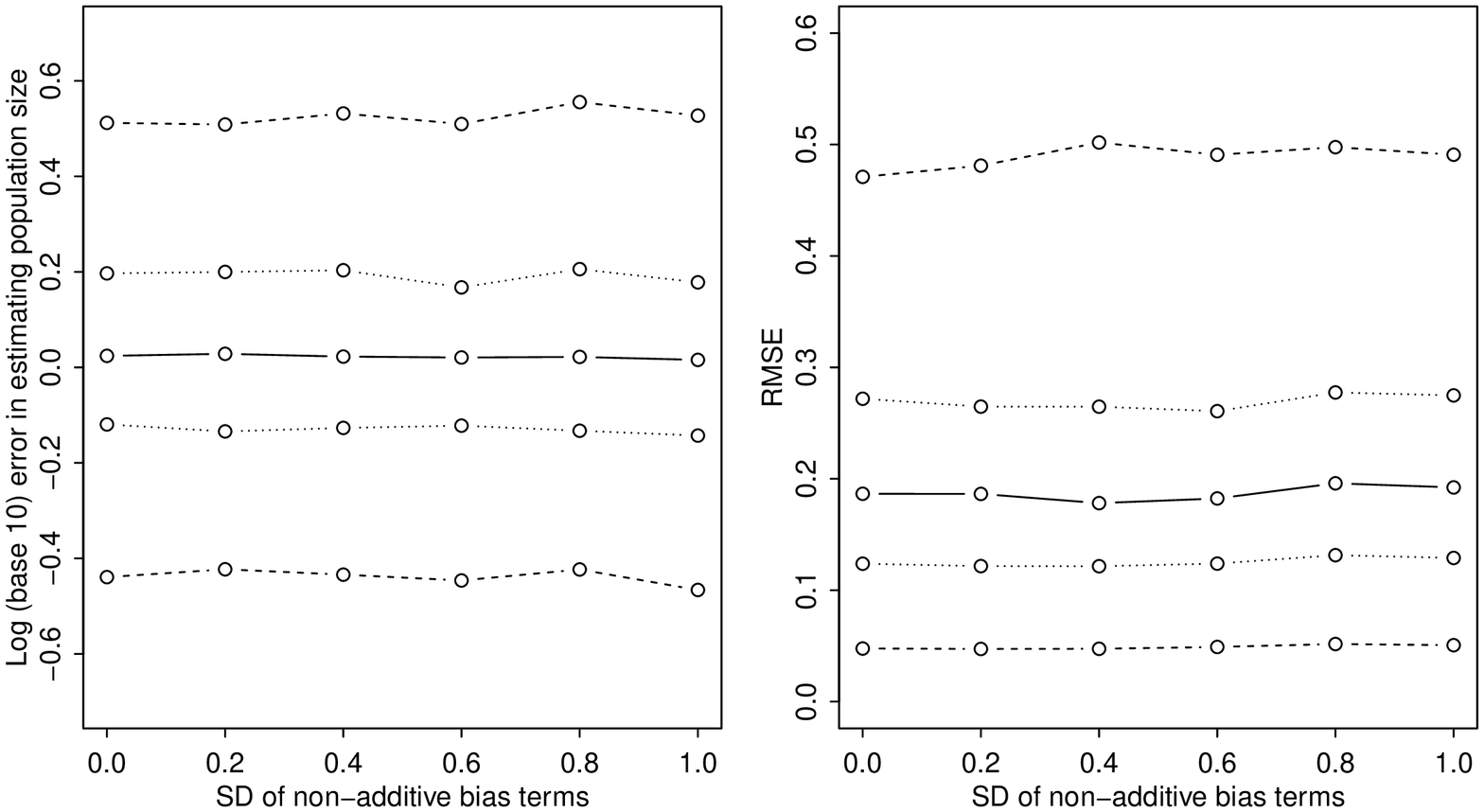}
	\caption{ The left panel shows the bias in estimating the population size for a city on the $log_{10}$ transformed scale plotted against the standard deviation $\sigma_c$ of the interaction terms $c_{ij}$. The solid middle line indicates the mean error in estimation across simulated data sets and cities, the middle dotted lines indicate bounds containing $50\%$ of the simulated average error across all cities and year combinations in each data set. The dashed lines indicate the middle $95\%$ errors in estimating population size for each city and year combination across all data sets. The right panel plots the mean root-mean square error in estimating the population size of all cities and years in each simulated data set with the solid line indicating the mean root mean square error, the dotted lines indicating the middle $50\%$ root mean square errors, and the dashed lines the middle $95\%$. }
		\label{fig:BiasInteraction}
\end{figure}

Figure \ref{fig:BiasInteraction} illustrates the distribution for the resulting errors in estimating the size of the target population at each city and time, the average error across all cities and time in a data set, and the root-mean square error for estimating all cities. No systematic trend in the errors is noticeable. In this setting we see that the non-additive bias terms have little impact on the mean estimates. 

This does not mean that violating this assumption has no effect on the mean estimates in any scenario.  If the bias terms are organized in a non-random way, then the mean estimates may end up being effected. For instance, if subgroups observed only in later years have interaction terms $c_{ij}$ that tend positive, we would be in a situation reflective of the previous section and have similar results. This sort of error could also be caused by simply having more positive $\gamma_j$ in later years reflecting the impact that missing data in a systematic manner can have on applications of this model.

\subsubsection{Temporal Trend Diagnostics}

\begin{figure}
	\centering
	\includegraphics[height=5cm]{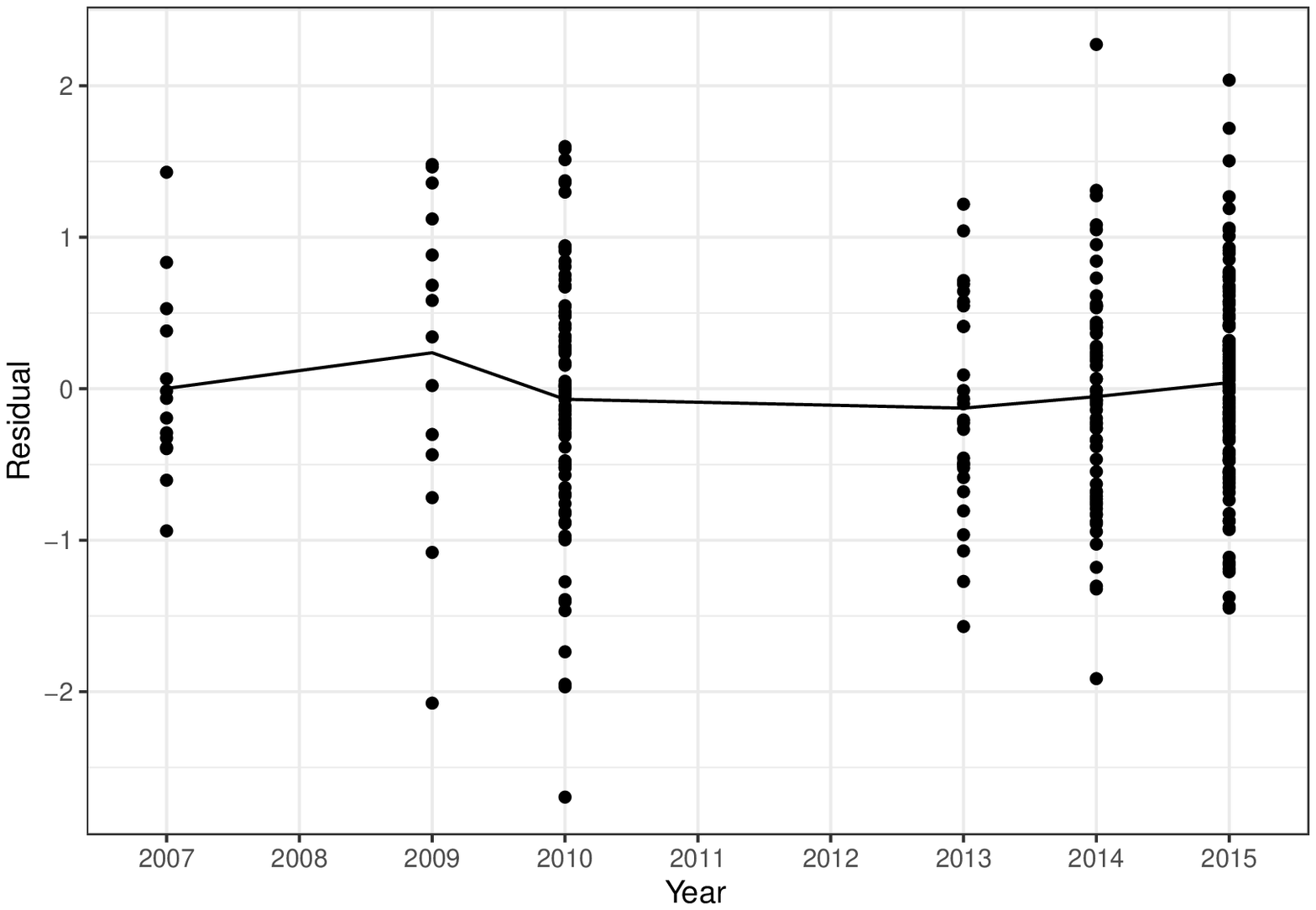}
	\includegraphics[height=5cm]{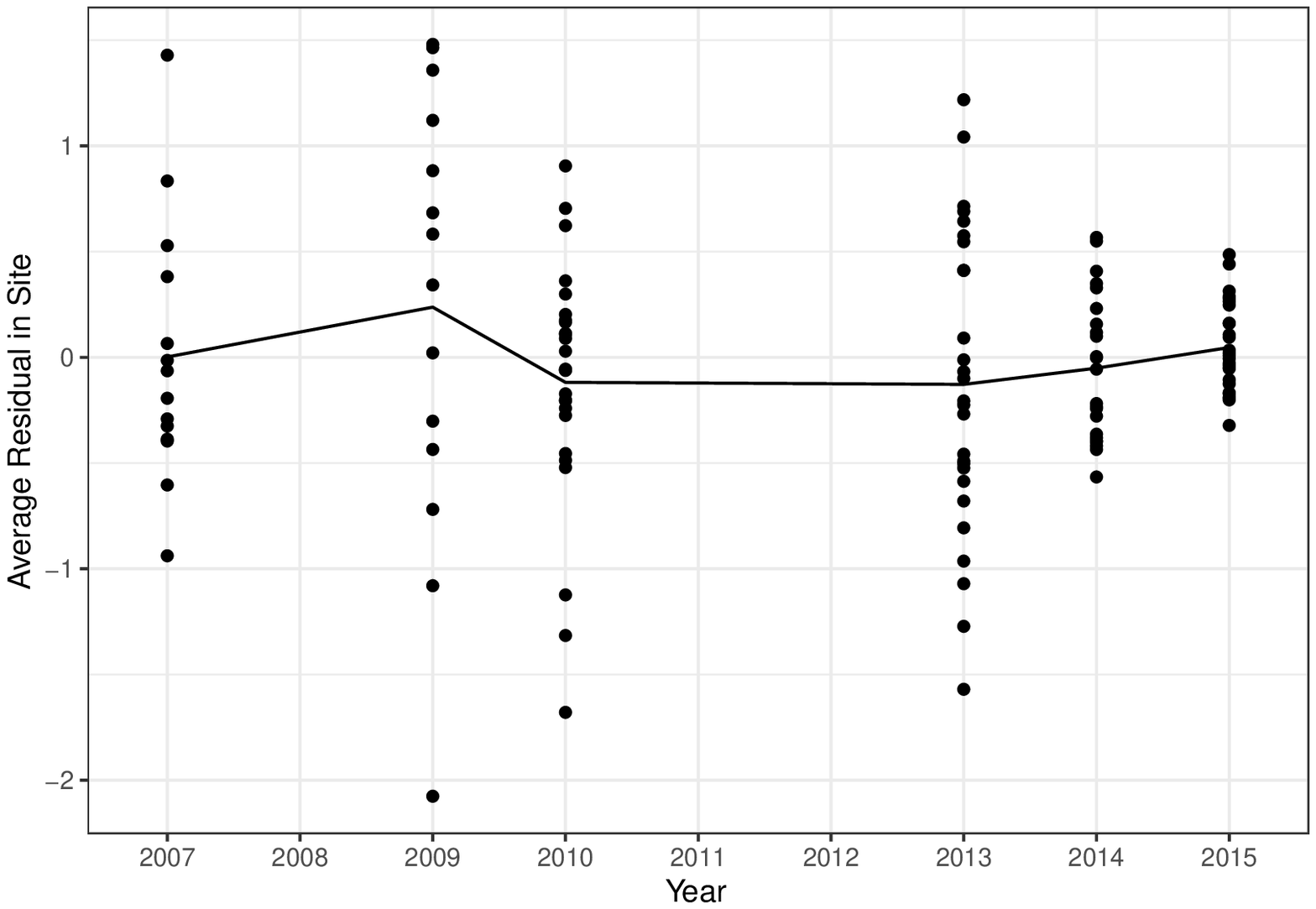}
	\caption{Left plot shows the residuals (difference between observed multiplier estimate and the expected multiplier estimate on the log scale) by year with a the line passing through the mean residual for each year. The right plot shows the mean residual within a site and year combination against the year. Wider bands correspond to years for which less supgroups  are available. }
	\label{fig:yearResid}
\end{figure}

We examine the assumption we've made for the temporal trend by investigating whether there is any obvious pattern or lack of fit in the residuals. Figure \ref{fig:yearResid} shows the residuals for the multiplier method portion of the model plotted against the year. The average deviation from the fitted value is represented by a line. The residuals have mean values close to zero for each year. All average deviations are within reasonable ranges with no obvious temporal patterns shown. This supports that the simple random walk assumption is reasonable.

\subsection{Evaluation of the distributional assumptions and goodness-of-fit}

\subsubsection{Distributional assumptions}
\begin{figure}[!ht]
	\centering
	\includegraphics[height=8cm]{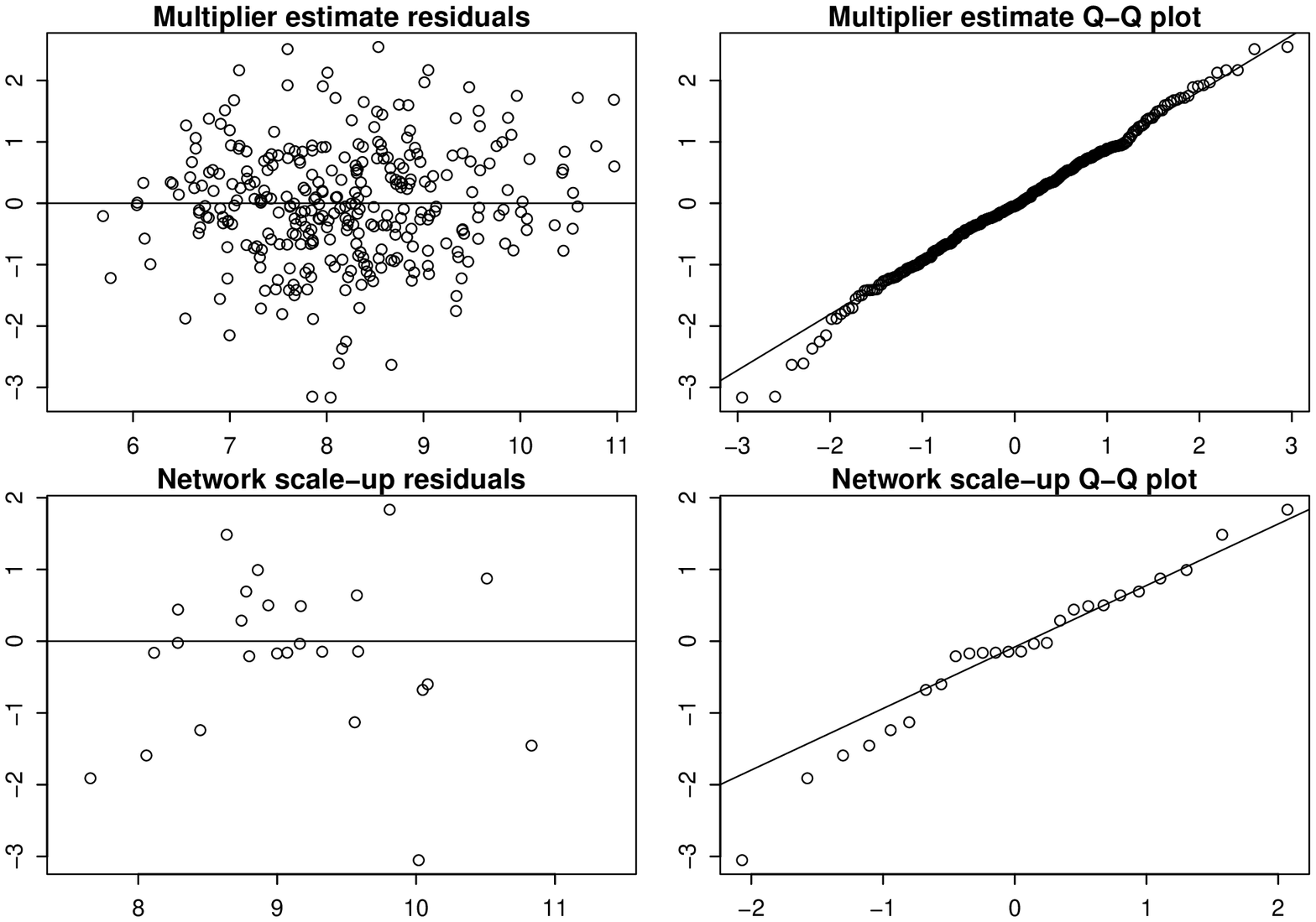}
	\caption{Residual plots for the model with residuals scaled so that they are expected to have a standard normal distribution under mean fitted values. The top row of plots shows the residuals for the multiplier method size estimates and the bottom row shows the residuals for the network scale-up size estimates. The left column shows the residuals plotted against the fitted values for each estimate and the right column plots the observed quantile of each observation against the theoretical quantile for each observation. }
	\label{fig:residuals}
\end{figure}

The distributions of the data being used can be checked for compatibility with the model by inspection of residuals. We inspect the behavior of the residual values based on posterior mean values of the parameters. We begin by considering the estimates for the proportion of people who inject drugs in the subgroups used in the multiplier method. Conditioning on all of the parameters the scaled errors$\sqrt{G_{ijt}} [\text{log}(Y_{ijt} / P_{ijt}) - \text{log}(n_{it}) - \theta - \delta_i - \gamma_j ]$
should be independently drawn from a zero mean normal distribution. As can be seen from the quantile-quantile plot in Figure \ref{fig:residuals} (top row) the distribution of the scaled residuals obtained by replacing the parameters with their posterior means is well approximated by a normal distribution. It can also be seen that the residuals appear to be centered at zero with no discernible relation between the fitted values and the residuals. That is, there is no clear systematic deviation from the model predictions. Similar plots comparing the scaled residuals against the population of the city, the sample size of the survey used to generate $P_{ijt}$, fits of individual parameters, and year also show no obvious patterns in the residuals.

Next, we move our attention to the network scale-up estimates. If we condition on all of the parameters, the scaled errors $\frac{S_{it}}{N_{it}}[log(N_{it}) - log(n_{it} )- \mu ]$
should be independent draws from a common normal distribution. A normal quantile-quantile plot of the residuals scaled so that they would be expected to be normal using mean posterior values for the parameters is shown in Figure \ref{fig:residuals} (bottom row). It can be seen that the normal distribution approximates the distribution of the estimates within reason given the sample size. 

\subsubsection{Prediction via cross-validation}
While the model seems to explain the behavior of the data it used for fitting the model, it is also of interest to see how well the model predicts data that it has not yet seen. To this end, we apply a leave one site out cross validation procedure. The prediction performance will be assessed visually, by the average coverage of a 95\% credible intervals for multiplier estimates and network scale-up estimates, and by the correlation between observations and predictions. 

\begin{figure}
	\centering
	\includegraphics[height=6cm]{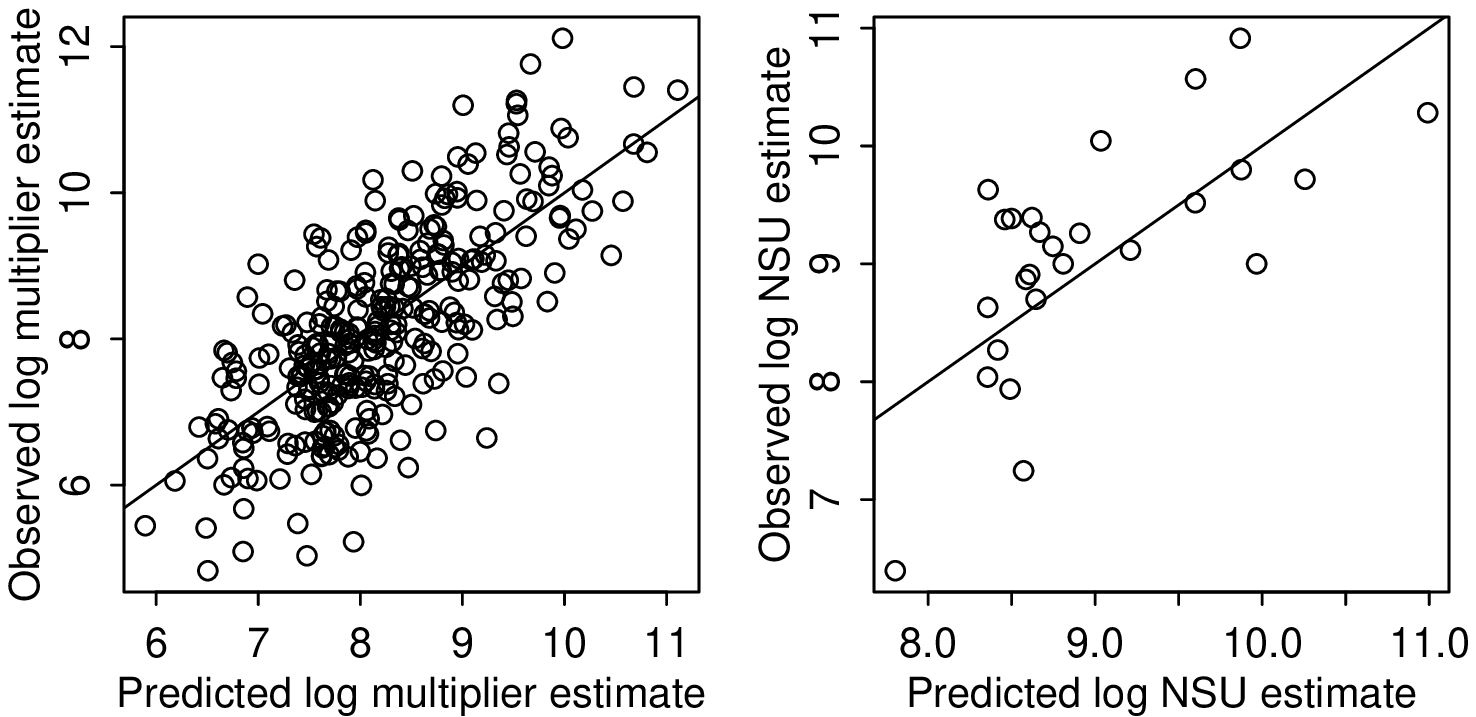}
	\caption{Predicted values for size estimates using the multiplier method and the network scale-up method based on the data with the city corresponding to that estimate removed plotted against the observed estimate.}
	\label{fig:predictions}
\end{figure}

The predicted values are positively correlated with the observed values for both sets of estimates as can be seen in Figure \ref{fig:predictions}.  The correlation between predicted and observed multiplier method size estimate on the log scale was $0.73$. The correlation between the predicted log scale network scale-up estimates and the observed estimates was $0.66$.  The fact that the observed estimates tend to vary largely about the predictions is expected. For the multiplier estimates, when the posterior distribution is computed without conditioning on a particular city, the prediction for the city is made based on the posterior distribution of the average proportion of individuals in each subgroup and year combination, the overall average bias $\theta$ of the multiplier estimates, and the bias for each subgroup $\gamma_j$. The spread corresponds to the variation due to the distribution of city bias $\delta_i$ and the variance of the estimator itself. For the network scale up data, the predicted values are based on the posterior distribution of the bias term $\mu$ and the size of the target population at a new city. 

For the multiplier estimates, the credible intervals are somewhat conservative with $99\%$ of the $95\%$ credible intervals for the removed estimates containing the true value for the multiplier estimate. For the network scale-up estimates, the 95\% credible intervals appear well calibrated with 96\% of the credible intervals containing the true value. 

\subsubsection{Posterior predictive checks}
\begin{figure}[!ht]
	\centering
	\includegraphics[scale=.7]{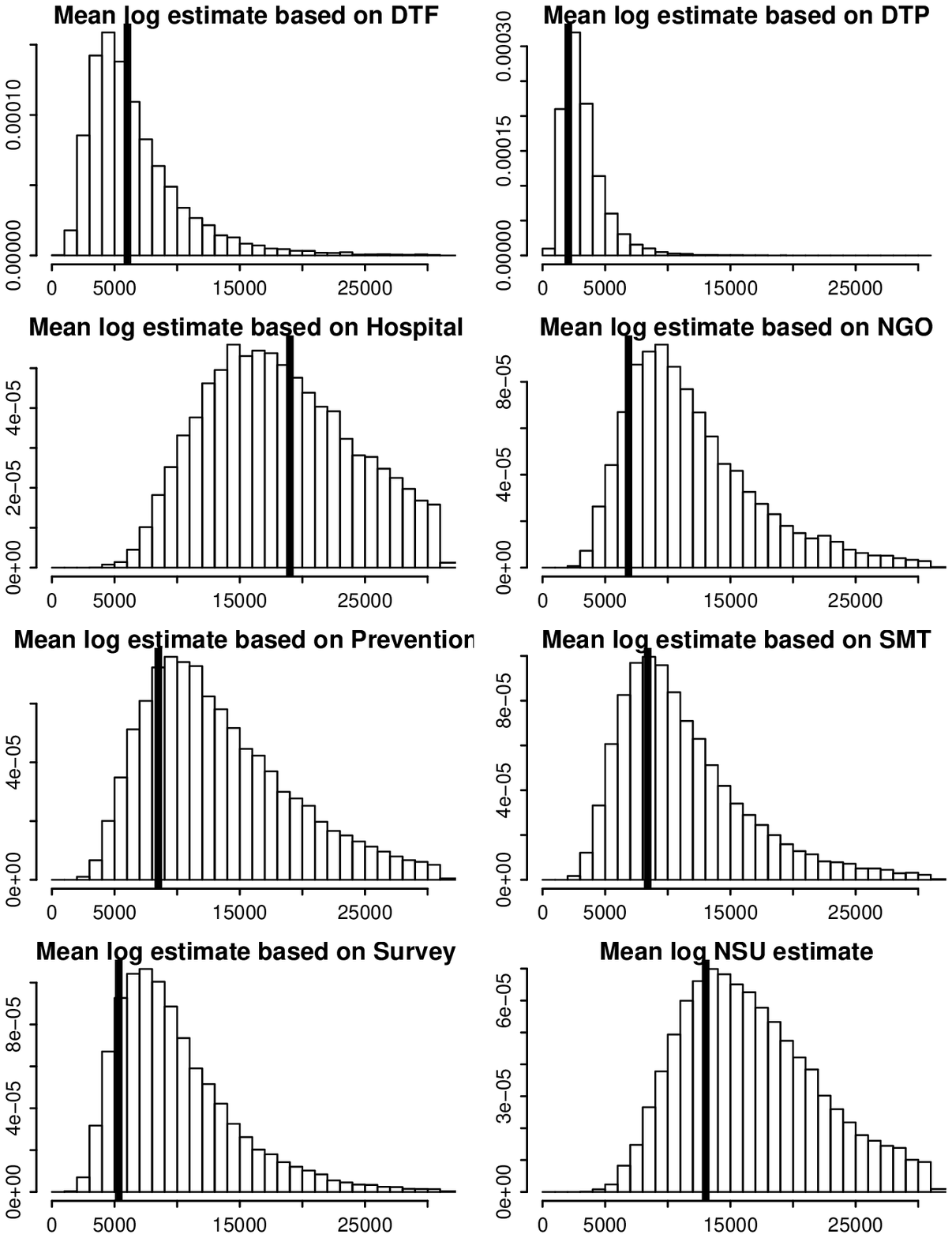}
	\caption{Histograms for the mean across all available cities for theoretical replications of the size estimate given by the multiplier method applied to each subgroup and given by the network scale-up estimates. The vertical lines correspond to the observed mean estimate across all cities for which a data source is available. }
	\label{fig:postpred}
\end{figure}

We use posterior predictive checks to examine how well the observed data correspond to theoretical replications of the data. We check the posterior predictive distributions of the mean of the multiplier estimates for each list and the mean of the network scale-up estimates across cities. As can be seen from Figure \ref{fig:postpred}, the observed values for each of these quantities is within the high probability regions for theoretical replications of these quantities under the posterior predictive distribution. Note the high variability predicted in the estimates reflecting the extreme variability in estimates observed even within the same city.